\documentclass[prd,nofootinbib,floats,superscriptaddress,eqsecnum,tightenlines,preprintnumbers,11pt]{revtex4-1}

\usepackage{hyperref}
\usepackage{graphicx}
\usepackage{amsmath,amssymb,amsfonts,amsthm,latexsym,stmaryrd}
\usepackage{marginnote}
\usepackage{xcolor}

\usepackage{physics}
\usepackage{csquotes}
\usepackage{tensor}
\usepackage{mathtools}
\usepackage{amssymb}
\usepackage{siunitx}
\usepackage{caption}
\usepackage{cleveref}
\usepackage{multirow}
\usepackage{subcaption}
\usepackage{rotating}
\usepackage[toc,page]{appendix}
\usepackage{nicefrac}

\def\be{\begin{equation}}
\def\ee{\end{equation}}
\def\beq{\begin{equation}}
\def\eeq{\end{equation}}
\newcommand{\bea}{\begin{eqnarray}}
\newcommand{\eea}{\end{eqnarray}}
\def\bi{\begin{itemize}}
\def\ei{\end{itemize}}
\def\ba{\begin{array}}
\def\ea{\end{array}}
\def\bfig{\begin{figure}}
\def\efig{\end{figure}}

\newcommand\rs{r_{\rm s}}
\newcommand\mass{r_{\rm m}}
\newcommand\mxi{\mu_\xi}

\newcommand\clam{\lambda}
\newcommand\ceta{\eta}
\newcommand\cU{\mathcal{U}}
\newcommand\cV{\mathcal{V}}

\newcommand\cS{{\mathcal{S}}}

\newcommand{\af}{f_0}
\newcommand{\bb}{f_1}
\newcommand{\gp}{{p_1}}
\newcommand{\rz}{r_-}
\newcommand{\rp}{r_+}
\newcommand{\rg}{r_g}

\newcommand{\Qa}{\Upsilon}
\newcommand{\rang}{r}
\newcommand{\ci}{a^\infty}
\newcommand{\ch}{a^{\rm hor}}
\newcommand{\co}{a}
\newcommand{\X}{Y}

\newcommand\cA{\Psi} 
\newcommand\cB{\Phi} 
\newcommand\cC{\Gamma} 
\newcommand\cD{\Delta} 
\newcommand\cF{\mathcal{F}}

\newcommand{\ca}{p}
\newcommand{\cb}{q}
\newcommand{\cc}{r}
\newcommand{\cd}{s}

\begin{document}

\title{
Black hole perturbations in modified gravity
}

\author{David Langlois}
\affiliation{Universit\'e de Paris,  CNRS, Astroparticule et Cosmologie, F-75013 Paris, France}
\author{Karim Noui}
\affiliation{Institut Denis Poisson (UMR 7013), Universit\'e de Tours, Universit\'e d'Orl\'eans, Parc de Grandmont, 37200 Tours, France}
\affiliation{Universit\'e de Paris,  CNRS, Astroparticule et Cosmologie, F-75013 Paris, France}
\author{Hugo Roussille}
\affiliation{Universit\'e de Paris,  CNRS, Astroparticule et Cosmologie, F-75013 Paris, France}
\affiliation{Institut Denis Poisson (UMR 7013), Universit\'e de Tours, Universit\'e d'Orl\'eans, Parc de Grandmont, 37200 Tours, France}

\date{\today}

\begin{abstract}
We study the linear perturbations about  nonrotating black holes in the context of degenerate higher-order scalar-tensor (DHOST) theories,  using a systematic approach that extracts the asymptotic behaviour of perturbations (at spatial infinity and near the horizon) directly from the first-order radial differential system governing these perturbations.
For axial (odd-parity) modes,  this provides an alternative to the traditional approach based on a second-order Schr\"odinger-like equation with an effective potential, which we also discuss for completeness. For polar (even-parity) modes, which contain an additional  degree of freedom in DHOST theories, and are thus more complex, we use a direct treatment of the four-dimensional first-order differential system (without resorting to a second order reformulation). We illustrate our study with two specific types of black hole solutions:  ``stealth'' Schwarzschild black holes, with a non trivial scalar hair,  as well as a class of non-stealth black holes whose metric is distinct from Schwarzschild. 
The knowledge of the asymptotic behaviours of the perturbations  enables  us to compute numerically quasi-normal modes, as we show explicitly for the non-stealth solutions.  Finally, the asymptotic form of the modes also signals   some pathologies in the stealth and non-stealth solutions considered here.
\end{abstract}

\maketitle

\section{Introduction}
The dawn of gravitational wave (GW) astronomy has spurred a renewed interest in possible deviations from General Relativity (GR), which could be detected in the GWs 
emitted by compact binaries.
Of particular interest is the ringdown phase of a binary black hole merger, which can be described by linear perturbations about a background stationary black hole solution. These perturbations  mainly correspond to a superposition of quasinormal modes, whose frequencies are quantised (see the  reviews \cite{Kokkotas:1999bd,Nollert:1999ji,Berti:2009kk,Konoplya:2011qq} and references therein). One expects that modified gravity models would predict QNMs that differ from their  GR counterpart and the detailed analysis of the GW signal, commonly called ``black hole spectroscopy", represents an invaluable window to test General Relativity and to look for specific signatures of modified gravity \cite{Berti:2005ys,Berti:2018vdi}. So far, QNMs  have been investigated only for a few models of modified gravity (see e.g.  the review  \cite{Berti:2018vdi} and references therein).

With these motivations in mind, the goal of this paper is to present a new  approach  for the study  of  black holes perturbations, illustrated  in the context of scalar-tensor theories, which constitute the simplest extensions of Einstein's theory. 
So far, the most general covariant scalar-tensor theories containing a single scalar degree of freedom are degenerate higher-order scalar-tensor (DHOST) theories, introduced and constructed   up to quadratic order (in second derivatives of the scalar field) in \cite{Langlois:2015cwa}  and extended  up to cubic terms  in \cite{BenAchour:2016fzp} (see  \cite{Langlois:2018dxi} for a review). DHOST theories encompass the traditional scalar-tensor theories (see e.g. \cite{Fujii:2003pa} and references therein), Horndeski's theories \cite{Horndeski:1974wa} and  Beyond Horndeski theories    such as the disformal transformations of GR \cite{Zumalacarregui:2013pma} and  GLPV theories \cite{Gleyzes:2014dya}.

There already exists a significant literature on black holes in DHOST theories 
\cite{BenAchour:2018dap,Motohashi:2019sen,Charmousis:2019vnf,Minamitsuji:2019shy,BenAchour:2019fdf,Minamitsuji:2019tet,Anson:2020trg, BenAchour:2020fgy,Takahashi:2020hso,Babichev:2020qpr,Baake:2020tgk} or in subclasses like Horndeski theories (see the review \cite{Babichev:2016rlq} and references therein) and Beyond Horndeski theories \cite{Babichev:2017guv}.
Among the solutions discussed in the literature, one can distinguish  the so-called stealth black holes, corresponding to solutions with a non-trivial scalar field  profile but with a metric that exactly coincides with a GR black hole solution (possibly with a cosmological constant). These solutions have been scrutinised in detail as they could naturally be compatible with present observations, while leading to specific signatures,  at the level of  perturbations, that could be detected or constrained by observations.  It appears however  that  stealth solutions seem to suffer from strong coupling issues or instabilities \cite{Minamitsuji:2018vuw,deRham:2019gha,Khoury:2020aya}. As suggested in \cite{Motohashi:2019ymr}, one possible cure to the strong coupling problem could be  a small  detuning of the degeneracy condition.

Other black hole solutions, distinct from GR solutions, have also been constructed. {Here} we will mainly consider a family of solutions introduced in \cite{Babichev:2017guv}, whose metric is formally analogous to that of Reissner-Nordstr\"om black hole but with the square of the electric charge effectively {\it negative}, which implies that there is a single horizon. In addition, the scalar field has a non trivial profile in this geometry.

 Beyond the construction of exact solutions, the linear  perturbations of  nonrotating black holes in DHOST theories, or in some subclasses, have been studied in a few papers. For background solutions in  Horndeski theories with a purely radially dependent scalar field,  the axial perturbations  were investigated in \cite{Kobayashi:2012kh} and the polar perturbations  in \cite{Kobayashi:2014wsa},  in both cases
by  reducing the quadratic action to keep only the physical degrees of freedom. This analysis was extended in \cite{Ogawa:2015pea,Takahashi:2016dnv} to include a linear time dependence of the background scalar profile, although the stability issue  was subsequently revisited  in \cite{Babichev:2018uiw}. Black hole perturbations were further discussed in \cite{Takahashi:2019oxz,Tomikawa:2021pca,Takahashi:2021bml}  in the context of DHOST theories. The perturbations of stealth black holes in  some  DHOST theories  were also investigated  in  \cite{deRham:2019gha}
and \cite{Khoury:2020aya},  showing that the equation of motion for the (polar) scalar degree of freedom is characterised by a singular effective metric in some cases, or concluding to the existence of a gradient instability in other cases.  The perturbations of the stealth Kerr black hole solutions found in \cite{Charmousis:2019vnf}  were analysed in \cite{Charmousis:2019fre}.

 Perturbations of nonrotating black holes in Horndeski theories were also  studied in 
\cite{Tattersall:2017erk}, but in the restrictive case of a constant background scalar field, which excludes the stealth and non-stealth black holes  with a non trivial scalar field profile. In this simple case, axial modes satisfy exactly the same equations as in GR,  while the equations of motion for the polar perturbations can be rewritten in a matricial Schr\"odinger-like system. The latter belongs  to the  family of generalised  second-order Schr\"odinger-like matricial systems  considered in  \cite{McManus:2019ulj} to parametrise small deviations from GR and  compute  the perturbations of quasi-normal frequencies with respect to their GR values. 

In the present work, instead of using a  second-order system, which in general is more complex that the ansatz considered in \cite{McManus:2019ulj} and requires a convoluted calculation (as illustrated in  \cite{Takahashi:2021bml} for stealth black holes), 
   we resort here to the new approach that we have presented in a companion paper \cite{Langlois:2021xzq}, to be referred to as Paper I. This method 
analyses directly the first-order differential system in its original form and extracts the asymptotic behaviour of the perturbations.
This enables us to identify the asymptotic behaviour of the physical modes and, in particular,   to  estimate numerically the quasi-normal modes, which are defined by their asymptotic boundary conditions. 
In this way, we are able to  get new insights concerning the perturbations of stealth black holes   and to explore for the first time the perturbations of a non-stealth solution introduced in \cite{Babichev:2017guv}.

The structure of the paper is the following. In the next section, we present the quadratic DHOST theories and the black hole solutions considered in the rest of the paper. In section \ref{section_axial_standard},  following the standard method,  we  write a general Schr\"odinger-like equation for axial perturbations, which is then applied to  our specific cases of interest. In section \ref{section_axial_new},  we revisit the axial perturbations with our novel approach, obtaining the asymptotic behaviours of the modes and computing numerically the QNMs for the non-stealth solution. 
We then turn, in section \ref{section_polar_new}, to the case of polar modes, for which the standard method is not available. We conclude in section \ref{section_conclusion}. Several appendices have also been added to provide more details on a few technical points.

\section{Black holes in DHOST theories}
\label{sec1}
 In this section, we give  a brief summary of quadratic DHOST theories, focussing on the  subclass Ia (according to the classification of \cite{Achour:2016rkg}) which contains the most interesting theories from a  phenomenological point of view. We then review  a few  static and spherically symmetric black hole solutions in these theories.

\subsection{Quadratic DHOST theories}
Allowing for second-order derivatives in the action, the most general family of viable scalar-tensor theories, which contain a single scalar degree of freedom and are  free from  Ostrogradski instabilities, can be constructed in a systematic way by requiring the degeneracy of the theories~\cite{Langlois:2015cwa}. 
Quadratic DHOST theories are described by an action of the form 
\begin{equation}
\label{DHOST}
S=\int d^4x\sqrt{-g}\left[P(X,\phi)+Q(X,\phi)\, \Box \phi+F(X,\phi)\,{}^{(4)}\! R+\sum_{i=1}^{5}A_{i}(X,\phi)\, L_{i}\right]
\end{equation}
where ${}^{(4)}\! R$ is  the Ricci scalar for the metric $g_{\mu\nu}$ and  the  $L_{i}$ denote the five possible scalar terms quadratic in second derivatives of $\phi$, namely 
\begin{eqnarray}
\label{L_i}
&&L_1 \equiv \phi_{\mu\nu} \phi^{\mu\nu} \, , \quad
L_2 \equiv (\Box \phi)^2 \, , \quad
L_3 \equiv \phi^\mu \phi_{\mu\nu} \phi^\nu\Box \phi \, , \quad \nonumber \\
&&L_4 \equiv  \phi^\mu  \phi_{\mu\nu} \phi^{\nu\rho} \phi_\rho \, , \quad
L_5 \equiv (\phi^\mu \phi_{\mu\nu} \phi^\nu)^2 \, ,
\end{eqnarray}
using the short-hand notations $\phi_\mu \equiv \nabla_\mu \phi$ and $\phi_{\mu\nu} \equiv \nabla_\nu \nabla_\mu \phi$ for the first and second (covariant) derivatives of $\phi$.
The action contains eight  functions,  $A_{i},\,F,\,Q$ and $P$, which depend on the scalar
field $\phi$ and its kinetic term $X\equiv \phi_\mu \phi^\mu$. 
While the functions $P$ and $Q$ are arbitrary, the functions $F$ and $A_i$  must satisfy three algebraic conditions \cite{Langlois:2015cwa}, in order to ensure the degeneracy of the theory and the absence of any Ostrogradski ghost.

As shown in \cite{Achour:2016rkg,Crisostomi:2016czh}, quadratic DHOST  theories  can be classified  into several  classes and subclasses which are 
stable under general disformal transformations, i.e. transformations of the metric of the form
\beq
\label{disf}
g_{\mu\nu} \longrightarrow \tilde g_{\mu\nu}= C(X, \phi) g_{\mu\nu}+ D(X,\phi) \phi_\mu \, \phi_\nu\,,
\eeq
where $C$ and $D$ are arbitrary functions such that the two metrics {$g_{\mu\nu}$ and $\tilde g_{\mu\nu}$} are not degenerate. Note that, when the disformal
transformation is not invertible, { one gets mimetic theories of gravity \cite{Chamseddine:2013kea,Deruelle:2014zza}, which can also be seen as DHOST theories \cite{Takahashi:2017pje,Langlois:2018jdg}. As shown recently in \cite{Langlois:2020xbc}, invertible disformal transformations can also be used to exhibit a remarkably simple Lagrangian for quadratic DHOST theories when ignoring matter.}

{Similarly},  the theories belonging to  class Ia can be mapped into a Horndeski form by applying a disformal transformation. 
The other classes are not physically viable (either tensor modes have pathological behaviour \cite{Langlois:2015skt} or gradient instabilities of cosmological perturbations are present
\cite{Langlois:2017mxy})
and will not be considered in the present work.
Theories in class Ia  are specified by the three free functions $F,A_1$ and $A_3$ (in addition to $P$ and $Q$) and 
the three remaining functions $A_2$, $A_4$ and $A_5$ are given by algebraic relations in terms of $A_1$, $A_3$, $F$ and $F_X$ (which denotes the derivative of  $F(X,\phi)$ with respect to $X$).
 These relations are a direct consequence of the three degenerate conditions, necessary to guarantee that  only one scalar degree of freedom is present \cite{Langlois:2015cwa,Langlois:2015skt}. In summary, this means that all the DHOST theories we consider here are characterized by the five functions $P$, $Q$, $F$, $A_1$ and $A_3$.

Finally, matter  can easily be  included by adding to the DHOST action an action $S_m$ where the matter degrees of freedom are minimally coupled to the metric $g_{\mu\nu}$, which therefore corresponds to the physical metric. Note that this implies that two DHOST theories that are disformally related via (\ref{disf}) are physically inequivalent when matter is included (assuming matter minimally coupled to $g_{\mu\nu}$ for the first theory and to  $\tilde g_{\mu\nu}$ for the second one).

If one is interested  only in {\it vacuum} solutions,  it can be  convenient  to use these disformal transformations to restrict the study of  DHOST Ia theories to their Horndeski  subclass, defined by the action
\begin{equation}
\label{Horndeski}
S =  \int \dd[4]{x} \sqrt{-g} \left[ F(X,\phi)\,  {}^{(4)}\! R + P(X,\phi) + Q(X,\phi) \Box\phi + 2F_X(X,\phi)\left(\Box\phi\right)^2 - \tensor{\phi}{_\mu_\nu}\tensor{\phi}{^\mu^\nu})\right],
\end{equation}
i.e.  DHOST Ia theories (\ref{DHOST}) with the restrictions 
\bea
A_1=-A_2= 2 F_X \, , \qquad  A_3=A_4=A_5=0 \, .
\eea
For simplicity,  in the following, we will study nonrotating black holes in gravitational theories described by the above Horndeski action.

\subsection{Black hole solutions}
We now consider static spherically symmetric black hole solutions, 
i.e. with a metric of the form 
\beq
\label{metric}
\tensor{g}{_\mu_\nu} \dd\tensor{x}{^\mu} \dd\tensor{x}{^\nu} = -A(r) \dd{t}^2 + \frac{1}{B(r)} \dd{r}^2 + r^2 \left(\dd{\theta}^2 + \sin^2\theta \dd{\varphi}^2\right),\\
\eeq
{where $A$ and $B$ are functions of the radial coordinate $r$ only.} 
Although it seems natural to assume a  radially dependent scalar field, i.e. of the form $\phi=\phi(r)$,  it was  realised in \cite{Babichev:2013cya} that one can adopt the more general ansatz
\beq
\label{phi_q}
\phi(t,r) = qt + \psi(r)\,,
\eeq
where $q$ is constant, in the context of {\it shift-symmetric} theories, i.e. where the arbitrary functions {entering in the DHOST action \eqref{DHOST}} depend only on $X$ and not on $\phi$. In this case, since only the gradient of the scalar field $\phi_\mu$ is relevant, (\ref{phi_q}) is compatible  with a static metric. 
Note that if $q\neq 0$ the disformal transformation of the metric \eqref{metric} does not conserve the same form, because  of the presence of a nonzero $\tilde{g}_{tr}\neq 0$. This implies that, in the case $q\neq 0$, working with the Horndeski action is more restrictive than starting with the general DHOST action.

 Even though our approach is  general,  in the following we will mainly concentrate   on two families of  interesting solutions  found in the literature, which we now introduce.

\subsubsection{Stealth solutions}
\label{stealthsolution}
Stealth solutions are solutions for which the metric coincides with a vacuum solution of General Relativity, possibly with a cosmological constant. This means that,  even if the scalar field profile is non trivial, i.e.  $\phi$ non constant, its effective energy-momentum tensor reduces to that of a cosmological constant.  These solutions have been actively studied in the context of Horndeski, beyond Horndeski and more generaly DHOST theories in the last few years
\cite{Babichev:2016rlq,Babichev:2017guv,Lehebel:2017fag,BenAchour:2018dap,Motohashi:2019sen,Takahashi:2019oxz,Minamitsuji:2019shy,BenAchour:2019fdf,Minamitsuji:2019tet}\footnote{Note  that stealth solutions were  first introduced in the context of three-dimensional gravity \cite{AyonBeato:2004ig} and an earlier stealth solution in four-dimensional modified gravity was discovered \cite{Mukohyama:2005rw}  in the context of ghost condensate~\cite{ArkaniHamed:2003uy} {(even though it was not named ``stealth'')}.  }.

For shift-symmetric DHOST Ia theories, or more specifically Horndeski theories, one can obtain stealth  Schwarzschild solutions with a scalar field satisfying  \eqref{phi_q} if the  conditions 
\beq
\label{stealthconditions}
X(x^\mu)=X_0 =- q^2\,, \qquad P(X_0)=P_X(X_0)=Q_X(X_0)=0\,,
\eeq
are satisfied\footnote{More general conditions to get stealth Schwarzschild solutions in DHOST Ia theories are \eqref{stealth_conds} and \eqref{stealth_conds2}.}. 
More concretely, the equations of motion involve the functions $F$, $P$ and $Q$ up to their second derivatives only evaluated at the background value $X_0=-q^2$, as can be seen in Appendix \ref{Appendix:EoM}.

As a consequence, if we fix $F(X_0)=1$ for convenience,  the only theory-dependent parameters  that appear in the equations of motion are
\bea
\label{paramHorn}
\alpha \equiv F_X(X_0) \, , \quad
\beta \equiv F_{XX}(X_0) \, , \quad
\gamma \equiv P_{XX}(X_0) \, , \quad
\delta \equiv Q_{XX}(X_0) \, .
\eea
In other words, without loss of generality, we can limit our study to Horndeski theories with 
\bea
&&F(X)\equiv 1 + \alpha (X+q^2) + \frac{\beta}{2}(X+q^2)^2 \, , \nonumber \\
&&P(X) \equiv  \frac{\gamma}{2}(X+q^2)^2 \, , \quad Q(X) \equiv  \frac{\delta}{2}(X+q^2)^2 \, .
\eea
All the other terms in the expansions in powers of $(X+q^2)$ of these functions  are irrelevant. 

\medskip

The stealth  Schwarzschild solution is then described by  the metric (\ref{metric}) with
\beq
\label{sch}
A(r) = B(r) = 1 - \frac{\rs}{r} \, ,
\eeq
where  $\rs$ denotes the  Schwarzschild radius, and the   scalar field  \eqref{phi_q} with\footnote{Note that the equations of motion lead to 
$\psi'$ up to a global sign.  Here we make one choice because it gives a regular expression (in Eddington-Finkelstein coordinates) while the  expression with the opposite sign leads to  a singular  scalar field on the horizon \cite{Babichev:2013cya}. However, such a  singularity has no physical consequences because $X$ itself and the stress-tensor energy are not singular.} 
 \beq
\label{stealth_psi}
\psi'(r) =  q \frac{\sqrt{ r \, \rs}}{r- \rs } \, ,  \qquad ({\rm stealth \  Schwarzschild})
\eeq
which is obtained by solving $X=-q^2$ (see \cite{Babichev:2013cya}). Throughout this paper, a  prime denotes a derivative with respect to the radial coordinate $r$.

\subsubsection{Babichev-Charmousis-Leh\'ebel (BCL) solutions}
\label{nonstealthsolution}
While it is natural to look for  stealth solutions in alternative theories of gravity, it is more interesting to find genuinely new solutions, i.e.  non-stealth solutions. For  DHOST theories,  this is  not an easy task as the equations of motion are quite involved,  even for a static and spherically symmetric metric. This is why very few exact  non-stealth solutions have been found so far\footnote{A new generic method to construct non-stealth solutions in DHOST theories has been introduced recently in \cite{BenAchour:2019fdf}. The idea  consists in using  a known solution $(g_{\mu\nu},\phi)$ of a given DHOST theory to build, via  a disformal transformation \eqref{disf},  a new solution $(\tilde g_{\mu\nu},\phi)$
for the disformally related DHOST theory. In general, a stealth solution transforms into a non-stealth one. An interesting result from this method is the construction of the first non-stealth rotating black hole solutions in DHOST theories \cite{Anson:2020trg, BenAchour:2020fgy}.}. Another approach is to construct  solutions numerically (see e.g. \cite{VanAelst:2019kku}  for rotating solutions in Horndeski theories with a cubic Galileon and a k-essence term only). 

As an illustration, we  study  in this work the  non-stealth  solutions  obtained in \cite{Babichev:2017guv} for a subset of  Horndeski theories \eqref{Horndeski} characterized by the functions 
\beq
F(X) = \af + \bb \sqrt{X},  \quad 
P(X) = -\gp X\,, \qquad Q(X) = 0\,,
\eeq
where $\af$, $\bb$ and $\gp$ are constants (we take $\af$, $\gp>0$) and $X$ is supposed to be positive. For simplicity, we  restrict ourselves to the case  where the scalar field \eqref{phi_q} has no time dependence, i.e. $q=0$.

The  black hole solution found in \cite{Babichev:2017guv}, which we will name BCL after the authors,    is described by a  metric of the form (\ref{metric}) with 
 \beq
\label{BCLrprm}
A(r) = B(r) =  \left(1-\frac{r_+}{r} \right)\left(1+\frac{r_-}{r} \right)\,,
\eeq
where $\rz$ and $\rp$ are defined by  the relations
\bea
\label{rprm}
r_+ r_- = \frac{\bb^2}{2\af \gp} \, , \qquad r_+ - r_- =\mass  \, \equiv  \, 2m \, , \qquad r_+>r_- > 0  \, .
\eea
Note that  the expression for $A(r)$ is reminiscent of the Reissner-Nordstr\"om metric but with a negative root here. As a consequence,  the black hole exhibits a single event horizon, of radius $r_+$, in contrast with the Reissner-Nordstroem geometry. 

As for the scalar field, its kinetic term  is given by 
\beq
\label{expressofXBCL}
X(r) =A(r)\phi^{\prime 2}(r) = \frac{\bb^2}{\gp^2 r^4},
\eeq
which is non constant, in contrast with  the stealth solutions presented above. The scalar field profile can be found explicitly by integrating the equation 
\bea
\phi'(r)  = \pm \frac{\bb}{\gp r \sqrt{(r-\rp)(r+ r_-)}} \, , 
\eea
yielding\footnote{The sign of $\phi(r)$ and the constant are physically  irrelevant. Notice that the derivative of the scalar field diverges at the horizon. According to \cite{Babichev:2017guv}, this is not a problem as it is a coordinate dependent statement which disappears in the tortoise coordinate for instance.  Furthermore, it was argued in \cite{Babichev:2017guv} that all physical meaningful quantities are well-defined at the horizon, for e.g.  the scalar field itself.}
\beq
\phi(r)=\pm \frac{\bb}{\gp\sqrt{ \rp\rz}}\arctan\left[\frac{\mass r +2\rp\rz}{2\sqrt{\rp\rz}\sqrt{(r-\rp)(r+\rz)}}\right]\, + {\text{cst}} \, .
\eeq
This concludes our presentation of the background solutions,  whose perturbations will be considered in the following.

\section{Axial perturbations: standard approach}
\label{section_axial_standard}
The rest of this paper is devoted to the  study of  the dynamics of linear perturbations about the black hole solutions described in the previous section. In this section and the next one, we examine  the axial (or odd-parity) perturbations, which are simpler  to analyse than  polar (or even-parity) perturbations discussed in section \ref{section_polar_new}. Axial perturbations correspond to the perturbations of the metric that transform like $(-1)^{\ell}$ under parity transformation, when decomposed into spherical harmonics, where $\ell$ is the usual multipole integer. 

In this section, we follow the standard approach for black hole perturbations which consists in reformulating the  linearised equations of motion as a   second order Schr\"odinger-like equation. In particular, we derive the corresponding effective potential for both stealth Schwarzschild and BCL black hole solutions.

\subsection{Equations of motion for the perturbations}
\label{Sec:eompert}
To derive the linearised equations of motion, let us  substitute  the perturbed metric and scalar field, 
\beq
\tensor{g}{_\mu_\nu}=\tensor{\overline{g}}{_\mu_\nu}+ \tensor{h}{_\mu_\nu} \,,\qquad \phi=\overline{\phi}+\delta\phi\,,
\eeq
where a bar denotes  a background quantity,  into the gravitational scalar-tensor  action \eqref{DHOST}, or \eqref{Horndeski}, and expand it  up to second order  in the perturbations 
$\tensor{h}{_\mu_\nu} $ and $\delta\phi$. The quadratic part of the action, $S_{\rm quad}[h_{\mu\nu},\delta \phi]$ then  describes the dynamics of linear perturbations and the linearised equations of motion are given by the associated   Euler-Lagrange equations,
\bea
{\cal E}_{\mu\nu}\equiv \frac{\delta S_{\rm quad}}{\delta h_{\mu\nu}}=0 \,,\qquad  {\cal E}_{\phi}\equiv \frac{\delta S_{\rm quad}}{\delta \phi}=0 \, .
\eea
The equation $ {\cal E}_{\phi}=0$ turns out to be redundant as a consequence of Bianchi's identities, so we just need to take into account the metric equations ${\cal E}_{\mu\nu}=0$.

\medskip
We now assume a background metric $\tensor{\overline{g}}{_\mu_\nu}$  of the form \eqref{metric}, keeping $A(r)$ and $B(r)$ unspecified at this stage, and a scalar field \eqref{phi_q}. 
In terms of the spherical harmonics $Y_{\ell m}$ and  working in the traditional  Regge-Wheeler gauge  (details about gauge fixing can be found in paper I), the axial metric perturbations for $\ell \geq 2$ read explicitly\footnote{As in GR, the dipole perturbation ($\ell=1$) does not not propagate \cite{Takahashi:2019oxz}.}
\begin{eqnarray}
&&h_{t\theta} = \frac{1}{\sin\theta}  \sum_{\ell, m} h_0^{\ell m}(t,r) \partial_{\varphi} {Y_{\ell m}}(\theta,\varphi), \qquad
h_{t\varphi} = - \sin\theta  \sum_{\ell, m} h_0^{\ell m}(t,r) \partial_{\theta} {Y_{\ell m}}(\theta,\varphi), \nonumber \\
&&h_{r\theta} =  \frac{1}{\sin\theta}  \sum_{\ell, m} h_1^{\ell m}(t,r)\partial_{\varphi}{Y_{\ell m}}(\theta,\varphi), \qquad
h_{r\varphi} = - \sin\theta \sum_{\ell, m} h_1^{\ell m}(t,r)  \partial_\theta {Y_{\ell m}}(\theta,\varphi), \label{eq:odd-pert}
\end{eqnarray}
while all the other components vanish.  Moreover, the scalar field  perturbation is zero by construction for axial modes.
All the modes  $(\ell m)$ decouple at the linear level and, in the following, we will drop this label to shorten the notation.

Since the metric is static, it is also convenient to decompose any time-dependent function $f(t,r)$  in Fourier modes, according to 
\begin{equation}
\label{eq:fourier}
f(t,r) = \int_{-\infty}^{+\infty} \dd\omega \, {f}(\omega, r)  \exp(-i \omega t) \, .
\end{equation}
In practice, this implies that all partial derivatives with respect to time become, in Fourier space, multiplications by $-i \omega$ .  The equations of motion for the Fourier modes, which we will also denote   ${\cal E}_{\mu\nu}=0$, therefore consist of a system of ordinary differential equations, with only derivatives with respect to the variable $r$.
As discussed in Appendix \ref{Appendix:EoM_perts_1}, the only relevant equations of motion reduce to
\begin{equation}
	\mathcal{E}_{r\theta} = 0, \quad \mathcal{E}_{\theta\theta} = 0 \,.
\end{equation}

These two equations are  first order ordinary differential equations and, after using the background equations of motion (see Appendix \ref{Appendix:EoM}), 
they  drastically  simplify  into a differential system for the two functions 
\beq
\label{X_axial}
\X_1(r)= h_0(r)\,, \qquad  \X_2(r) \equiv \frac{1}{\omega}\left(h_1+{\cA} \,  h_0\right)  \,,
\eeq
which reads, using $\X={}^T\!(\X_1,\X_2)$,
\bea
\label{generaloddsystem}
 \dv{{\X}}{r} = {M}(r) {\X} \, , \qquad 
 M \equiv \begin{pmatrix} {2}/{r}+i\omega {\cA} &- i \omega ^2 +2 i {\clam}{\cB}/{r^2}   \\ -i{ \cC} & {\cD} +i\omega\cA  \end{pmatrix} \, ,
\eea
where  
\beq
\clam \equiv \frac{(\ell-1)(\ell+2)}{2} \,,
\eeq
and we have introduced the four functions
\bea
&&\cA \equiv \frac{2 q F_X \psi'}{\cF} \,  , \quad \cB \equiv \frac{\cF}{F-2X F_X} \, ,\quad \cC \equiv \frac{AF(F-2X F_X)}{B {\cF}^2}  \, , \label{expressionsofABC}  \\ 
&&\cD \equiv \frac{\sqrt{A/B}}{\cF}  \left[ 
AB \left( \frac{F}{\sqrt{AB}}\right)' + 2X (\sqrt{AB} F_X)' + 2 q^2 \left( \sqrt{\frac{B}{A}} F_X\right)' 
\right] \, , \label{expressofD}
\eea
which depend on  the function $\cF$ defined by
\bea
&{\cF} \equiv A(F-2X F_X)-2q^2 F_X\, . 
\eea

In General Relativity, the  functions defined above reduce to
\beq
\label{expressofD_GR}
 \cA=0\,, \qquad \cB=A\,,\qquad  \cC= A^{-2}\,,\qquad  \cD=-A'/A\,,
\eeq
with $\cF=A$, 
and  \eqref{generaloddsystem} reduces to the system studied in Paper I.

\subsection{Schr\"odinger-like equation  and effective potential}
\label{Sec:HorndeskiSchro}

Following the standard  approach (originally introduced in \cite{Regge:1957td} and recalled  in Paper I), we now  recast the above system  \eqref{generaloddsystem} into a single Schr\"odinger-like equation, which is second order with respect to the radial coordinate $r$ and depends on $\omega^2$ (corresponding to a second order time derivative).

As discussed in detail in Appendix \ref{Sec:3DtoSchro}, a 
transformation
\beq
\label{change_functions}
\X(r)= \hat P(r) \hat\X(r)\,
\eeq
with the appropriate matrix\footnote{The matrix $\hat P$ here corresponds to the matrix $\tilde{P}\hat P$ in Appendix \ref{Sec:3DtoSchro}, where $\tilde{P}$ is defined in (\ref{App_Ptilde}) and $\hat P$ in (\ref{generalP}).}   $\hat P$
enables us to rewrite  the system  \eqref{generaloddsystem}   in the canonical form
\begin{equation}
\label{eq:eqs-matrix-odd}
 \dv{\hat \X}{r_*} = \begin{pmatrix} i\mu(r)\omega & 1 \\ V(r) - {\omega^2}/{c^2(r)} & i\mu(r) \omega\end{pmatrix}\hat \X \, ,
\end{equation}
where we have introduced a new radial coordinate $r_*$ and the functions $\mu$ and $c$, defined by
\bea
\dv{r}{r_*}  \equiv n \, , \qquad
\mu \equiv  n\,  {\cA} \, , \qquad c^2 \equiv  \frac{1}{n^2\,  {\cC}} \,.
\label{tortoiseHorn}
\eea
In terms of the functions introduced in \eqref{expressionsofABC}-\eqref{expressofD}, the potential $V$  in \eqref{eq:eqs-matrix-odd} reads (see Appendix \ref{Sec:3DtoSchro} for the explicit calculation)
\bea
\label{generalpotentialtext}
V  &= & \frac{n^2}{4} \left[ \frac{8(1+ \lambda \cB \cC) }{r^2}+ \cD^2 - \frac{4 \cD}{r} + 2 \cD' + 
\frac{2\cC'}{\cC} \left( \frac{2}{r} - \cD \right) \right. \nonumber \\
 &&\left. \qquad + 3 \left( \frac{\cC'}{\cC} \right)^2 + \left(\frac{n'}{n} \right)^2
- 2  \left(\frac{\cC''}{\cC} + \frac{n''}{n} \right)  \right]\, .
\eea
One can check that this coincides with the expression found in \cite{Takahashi:2019oxz}\footnote{See Eq. (64) of \cite{Takahashi:2019oxz}. We do not recover, however,  the potential used in  \cite{Ganguly:2017ort} in the limit $q\longrightarrow 0$. An unfortunate consequence is that the computation of quasi-normal modes in \cite{Tattersall:2019nmh} should be revisited as the latter potential was used in that work.}  in the case $n=\sqrt{AB}$.
 Let us stress that the explicit expressions of  propagation speed $c(r)$ and of the potential $V(r)$ depend on the choice of the radial coordinate $r_*$, characterised by $n$. 

In contrast to General Relativity, where $\cA=0$, one cannot eliminate in general the diagonal components in the new matrix above via a change of functions (\ref{change_functions}). However, as noticed in \cite{Takahashi:2019oxz}, this
 can be achieved via a time redefinition of the form\footnote{{Equivalently, one can get rid of the diagonal terms by a  redefinition  $\hat\X  \longrightarrow e^{-i \omega \nu(r)} \hat\X$, where $\nu(r)$ is given by \eqref{eq:choice-nu}. }}
\bea
\label{changetime}
t \longrightarrow t + \nu(r) \,.
\eea
Indeed, under such a time coordinate change, one easily show that the system \eqref{eq:eqs-matrix-odd}  transforms into
\begin{equation}
\label{eq:eqs-matrix-odd-nu}
\dv{\hat \X}{r_*}= \begin{pmatrix} i \omega (\mu(r) +  \dv*{\nu}{r_*})& 1 \\ V(r) - {\omega^2}/{c^2(r)} & i \omega (\mu(r) + \dv*{\nu}{r_*})\end{pmatrix} \hat \X \, .
\end{equation}
We can then eliminate the diagonal terms by choosing
\begin{equation}
\label{eq:choice-nu}
\nu(r) =- \int \frac{\mu(r)}{n(r)} \dd{r} 
= - \int \Psi(r) \dd{r} 
\end{equation}
This choice of change of time variable leads to a Schr\"odinger-like equation, of the form
\bea
\label{schroedinger}
\dv[2]{\hat{\X}_1}{r_*}  + \left(\frac{\omega^2}{c^2(r)} - V(r)\right) \hat{\X}_1 = 0 \, ,
\eea
where $c(r)$ corresponds to the propagation speed and $V(r)$ to the effective potential.

\subsection{Stealth Schwarzschild axial pertubations}
\label{Sec_Stealthpotential}

Let us apply the above results to  the stealth Schwarzschild solution described in \autoref{stealthsolution}. 
Substituting the background expressions  (\ref{sch}) and (\ref{stealth_psi}) into \eqref{expressionsofABC} and \eqref{expressofD}, one finds 
\beq
\begin{aligned}
\label{ABCStealth}
&
\cA = \frac{\zeta \, \rs^{1/2} r^{3/2} }{(r-\rs)(r-\rg)}\,, \quad 
\cB = \frac{r-\rg}{(1+\zeta)r} \,, \\
&\cC = \frac{(1+\zeta)r^2}{(r-\rg)^2}\,, \quad
\cD = \frac{1}{r} - \frac{1}{r-\rg} \,,
\end{aligned}
\eeq
where we have introduced  the constant parameters 
\bea
\label{czetadef}
\zeta \,  \equiv \, 2q^2\alpha\geq 0 \,,\qquad \rg\equiv (1+\zeta)\rs\,.
\eea
This dimensionless constant $\zeta$ parametrises the deviation from General Relativity, since one recovers the GR functions  \eqref{expressofD_GR} when $\zeta=0$. 
 The radius $\rg$,  which differs from $\rs$ when $\zeta\neq 0$, appears as an extra pole in the above functions, in addition to $\rs$ and $0$.

\medskip
From the expressions \eqref{ABCStealth}, one can compute the potential $V(r)$ and the propagation speed  $c(r)$ that appear in the Schr\"odinger-like equation \eqref{schroedinger}. 
 As already stressed in the previous subsection, these quantities depend on the choice of the radial coordinate. If one adopts the usual Schwarzschild tortoise coordinate, defined by
\bea
\label{TortoiseS}
r_* = \int \dd{r} \, \frac{r}{r- {\rs}} = r + \rs \ln (r/\rs-1) \,,
\eea
corresponding to the choice $n=A(r)=1- \rs/r$, 
the potential takes the form
\bea
\label{SSSpot}
V(r) = \frac{V_0 +V_1 \, (\rs/r) +  V_2 \,(\rs/r)^2 +V_3\, (\rs/r)^3 + V_4 \, (\rs/r)^4 }{ (r-\rg)^2} \, , 
\eea
with
\bea
&&V_0 = 2 (\lambda +1) \, , \quad V_1 = -2   (\lambda +3)\zeta-6 \lambda -9 \, , 
\quad V_2 =  (15 \zeta +16 \lambda +70) \zeta/4+6 \lambda +15 \, , \cr
&&V_3 = -(1+ \zeta) (13 \zeta/2 +2 \lambda +11) \, , \quad V_4 = 3 (1+ \zeta )^2  \,,
\eea
and the propagation speed is given by the expression
\beq
c(r)= \frac{r-{\rg}}{\sqrt{1+\zeta}\, ( r-{\rs})}\,,
\eeq
where  one must take  $\zeta >-  1$ in order to have $c^2>0$.

Another possibility is to choose the radial coordinate such that the propagation speed is $c=1$, i.e. 
\beq
\label{rstar_stealth}
r_* = \int \dd{r} \,\sqrt{\Gamma}= \sqrt{1+\zeta}\left[\ r + \rg \ln (r/\rg-1)\right] \, ,
\eeq
which is very  similar to the usual tortoise coordinate, with $\rg$ instead of $\rs$ and a global rescaling. In this case, the potential becomes 
\beq
\label{potc=1stealth}
V_{c=1}(r)= \left(1 - \frac{\rg}{r}\right) \frac{2 (\lambda +1)r - 3\rg}{(1+\zeta)\,  r^3}\,,
\eeq
which is, quite remarkably, identical to the standard Regge-Wheeler potential, with $\rg$ instead of $\rs$, up to  a global rescaling. One can note that $\rs$ has completely disappeared from the equation of motion and $\rg$ seems to play the role of the horizon that is effectively ``seen" by the axial metric perturbations. The same result was obtained recently in 
\cite{Tomikawa:2021pca} by analysing the effective metric that appears in the equation of motion for the axial perturbations. 

In fact, this result can be understood by noting that the quadratic Lagrangian for the axial tensor perturbations can come only from two terms in (\ref{DHOST}):  the  term in ${}^{(4)}\! R$, which contains both $K_{\mu\nu}K^{\mu\nu}$ and ${}^{(3)}\! R$ according to the Gauss-Codazzi identity\footnote{In a $(3+1)$ decomposition of spacetime, where $n^\mu$ is the unit vector normal to the spatial hypersurfaces, the Gauss-Codazzi equation reads
${}^{(4)}\! R  =   K_{\mu\nu} K^{\mu\nu} - K^2+ {}^{(3)}\! R + 2 \nabla_\mu (K n^\mu-n^\nu \nabla_\nu n^\mu ) $, where $h_{\mu\nu} \equiv g_{\mu\nu} + n_\mu n_\nu$ and  $K_{\mu\nu} \equiv  h_\mu^\alpha \, \nabla_\alpha n_\nu$.}, and the Lagrangian $L_1$ in \eqref{L_i}, which contains $K_{\mu\nu}K^{\mu\nu}$
(see discussion in \cite{Langlois:2017dyl}).  These two terms give, in the quadratic Lagrangian of the axial modes, a kinetic term  with coefficient  $F-XA_1$, evaluated on the background, and a gradient terms  with coefficient given by $F$,  again evaluated on the background. Since $X$ is a constant, these coefficients are constant and, when $A_1=0$,  one recovers the same quadratic Lagrangian as in GR  with $c=1$. Even if $A_1\neq 0$ (which is the case here since $A_1=2F_X=2\alpha$),  it is possible to perform a disformal transformation to go into a ``frame" where $A_1=0$ and therefore $c=1$. The background metric is disformally transformed into a new metric, which is straightforward to compute using the disformal transformations of quadratic DHOST theories given in  \cite{Achour:2016rkg}.
It turns out that this new metric is another stealth Schwarzschild metric with a displaced horizon, corresponding to $\rg$,  as discussed in \cite{BenAchour:2019fdf}, which explains why the potential in this frame coincides with the standard Regge-Wheeler potential.

\subsection{BCL axial pertubations}

We now apply the results of  \autoref{Sec:HorndeskiSchro} to the  non-stealth solution described in  \autoref{nonstealthsolution}.
 In this case, the  new coordinate $r_*$ is given by 
 \bea
\label{TortoiseBCL}
r_* = \int \dd{r} \, \frac{r^2}{(r - {r_+})(r +{r_-})} = r + \frac{r_+^2 \ln(r/r_+-1) - r_-^2 \ln(r/r_- + 1)}{r_+ +r_-} \, .
\eea

For the BCL  background, characterised by \eqref{BCLrprm} with \eqref{rprm}, we find  that the functions \eqref{expressionsofABC} and \eqref{expressofD} entering in the coefficients of the differential system \eqref{generaloddsystem} read
\bea
&& \cA = 0 \, , \qquad \cB = A \, , \qquad
 \cC = \frac{F}{\af A^2}=\frac{r^2(r^2 + 2 r_+ r_-)}{(r-r_+)^2 (r+r_-)^2} \, ,  \nonumber \\
&&
\cD = - \frac{A'}{A}= -\frac{r_+}{r(r-r_+)} +  \frac{r_-}{r(r+r_-)} \, , \label{coeffcalBCL}
\eea
since $\cF=\af A$. 

Furthermore,  the potential \eqref{generalpotentialtext} takes the form
\bea
\label{pot_bcl}
V(r) = A(r)\frac{  V_0 + V_1(\mass/r) + V_2 (\mass/r)^2+ V_4 (\mass/r)^4 +V_6 (\mass/r)^6 }{2 r^2 (1+ \xi (\mass/r)^2)^2}  \, ,
\eea
with  the coefficients
\bea
V_0=4(\lambda +1) \, , \quad
V_1=- 6 \, , \quad
V_2 =  6(2\lambda -1) \xi \, , \quad
 V_4=(12 \lambda -1) \xi^2 \, , \quad  V_6= 4 \lambda \xi^3 \,,
\eea
and where we have  introduced the dimensionless constant
\bea
\xi \equiv 2 \frac{r_+r_-}{\mass^2} \, = \, \frac{\bb^2}{ \af \gp \mass^2} \,.
\label{defofxi}
\eea
Similarly to the parameter $\zeta$ in the stealth case, $\xi$  parametrises the deviation from GR (corresponding to the limit $r_-=0$, i.e. $f_1=0$). 

\begin{figure}[!htb]
 \captionsetup{singlelinecheck = false, format= hang, justification=raggedright, font=footnotesize, labelsep=space}
	\centering
	\includegraphics[width=0.6\textwidth]{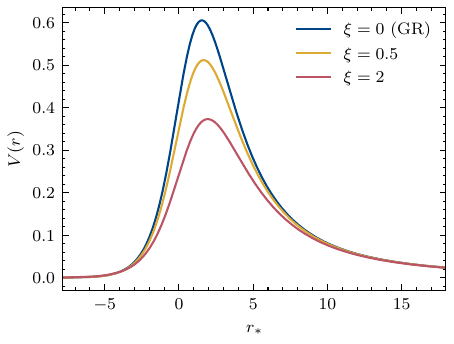}
	\caption{Potential $V(r)$ for the non-stealth for different values of $\xi$ but fixed values of $\mass=1$ and $\ell = 2$ ($\lambda=2$).}
	\label{fig:Veff_nonstealth}
\end{figure}

One notes that one must have $\xi\geq 0$ to prevent a singularity in the potential. 
When $\xi=0$, one recovers the standard Regge-Wheeler (RW) potential for the  Schwarzschild geometry,
\beq
\label{RWpotential}
V_{\rm RW}(r)= \left(1-\frac{\rs}{r}\right)\frac{2(\lambda+1) r-3\rs}{r^3}
\,,
\eeq
where $\rs=\mass$ in this limit. 
 Potentials for several values of $\xi$  are shown on Fig. \ref{fig:Veff_nonstealth}, where one can see that the potential is a deformation,  parametrised by $\xi$, of the  RW potential. At infinity, the behaviour of the potential is very similar to that of the RW potential, with corrections appearing only at second order in $\mass/r$:
 \beq
 V(r)=\frac{1}{\mass^2}\left[2(\lambda+1)\frac{\mass^2}{r^2}- (2\lambda+5)\frac{\mass^3}{r^3}+{\cal O}\left(\frac{\mass^4}{r^4}\right)\right]\,.
 \eeq
  By contrast, the leading order  behaviour is modified near the horizon,  
  \beq
  V(r)=\frac{32\mxi\left(\lambda(3 \mxi-1)^2-\mxi(1+\mxi)\right)}{(1+\mxi)^5(3\mxi-1) \, \mass^3}(r-\rp)+{\cal O}((r-\rp)^2)\,,\qquad \mxi\equiv \sqrt{1+2\xi}\,,
  \eeq
 where we have used $r_\pm = \mass (1\pm \mxi)/2$.  Notice that the height of the potential also depends on the value of $\xi$.

The  propagation speed is given by
\beq
\label{c_bcl}
c(r)=\frac{r}{\sqrt{r^2+\xi {\mass^2}}}\,.
\eeq
We thus recover the usual value $c=1$ at spatial infinity (when $r \rightarrow \infty$), but at the horizon 
we find 
\beq
\label{c_hor_bcl}
c(\rp)= \sqrt{\frac{r_+}{r_+ + 2{\rz}}}=\sqrt{\frac{\mxi+1}{3\mxi-1}} \leq 1 \, .
 \eeq

From the Schr\"odinger equation (\ref{schroedinger}) with (\ref{pot_bcl}) and (\ref{c_bcl}), one can compute explicitly the complex frequency of the associated quasi-norrmal models (QNM) by  resorting to standard numerical techniques
\cite{Kokkotas:1999bd,Nollert:1999ji,Berti:2009kk} which were applied in the context of Horndeski theories in \cite{Tattersall:2018nve}. In the present case, we will postpone the computation of the QNMs modes to the next section, where we will show that they can be computed numerically even without the Schr\"odinger-like reformulation of the equations of motion.

\section{Axial  perturbations: first-order system approach}
\label{section_axial_new}
In this section, we revisit  axial  perturbations and study their asymptotic behaviour, both at infinity and near the horizon, using the method presented in  Paper I.
Using these asymptotic behaviours, we  then compute the quasi-nomal modes  numerically.

\subsection{First order approach: method and goal}

Ignoring the traditional Schr\"odinger reformulation of the perturbation equations, discussed in the previous section, we now go back to the original first-order system and apply the systematic method developed in  Paper I   to determine the asymptotic behaviour of the solution at spatial infinity and near the black hole horizon. 

More precisely, let us consider some first order system of the form 
\bea
\label{system}
 \dv{{\X}}{z} =  M(z)  \X \, , \qquad 
 M(z) = z^\rang \sum_{n=0}^{p} M_{\rang-n} z^{-n} + {\cal O}(z^{\rang-p-1})\, ,
\eea
where $\X$ is a column vector, $M(z)$ a square matrix which can  expanded, up to some given order, when the variable $z$ goes to infinity. 
In most cases\footnote{The variable $z$ that appears in the asymptotic solution  \eqref{gensol} can sometimes differ from the original variable in the system \eqref{system}. Moreover, in the very particular cases where the system is  such that $M(z)=M_{-1}/z + {\cal O}(1/z^2)$ with $M_{-1}$ nilpotent, the asymptotic expansion of $\X(z)$ is no longer given by  \eqref{gensol} but it can be expressed as a polynomial of $\ln z$ (see section IV.C of Paper I).},
  the solution $\X(z)$ can be written asymptotically in the form 
\begin{equation}
\label{gensol}
  \X(z) =  e^{\mathbf{\Qa}(z)} \,  z^\mathbf{\Delta} \, \mathbf{F}(z)  \, \X_0, \qquad (z\rightarrow\infty)
\end{equation}
where $\X_0$ is a constant vector (which can be constrained by boundary conditions), $\mathbf{F}(z)$ is a matrix regular at infinity,  $\mathbf{\Delta}$ is a constant diagonal matrix, and  finally $\mathbf{\Qa}(z)$  is also a diagonal matrix whose coefficients are polynomials of degree (at most) $\rang$. 
The algorithm described in Paper I, based on \cite{wasow_asymptotic_1965,balser_computation_1999,barkatou_algorithm_1999,pflugel_root-free_2019}, 
enables one to compute explicitly all the quantities entering in \eqref{gensol}, up to some order. 

 Note that there is no loss of generality when considering the asymptotic behaviour at {\it infinity} since one can always reformulate a system that is singular  for some finite value $z_0$ into a system of the form (\ref{system}) via a change of variable.
 
 In the following, we apply the algorithm of paper I successively to the BCL perturbations and to the stealth Schwarzschild perturbations.

\subsection{BCL axial perturbations}

As  found in the previous section, the  axial perturbations of the BCL black hole   satisfy the system (see \eqref{generaloddsystem})
\bea
\label{systemBCLodd}
 \dv{{\X}}{r} = M \X \, , \qquad M(r) = \begin{pmatrix} {2}/{r} & -i \omega^2 + i 2\lambda {A}/{r^2}  \\ -i \cC &  \cD \end{pmatrix} \, ,
 \eea
 with, according to  \eqref{coeffcalBCL},
 \bea
 \label{ACDBCL}
 A=  \left(1-\frac{r_+}{r} \right)\left(1+\frac{r_-}{r} \right)\,, \quad
  \cC = \frac{r^2(r^2 + 2 r_+ r_-)}{(r-r_+)^2 (r+r_-)^2} \, , \quad
\cD = -\frac{r_+}{r(r-r_+)} +  \frac{r_-}{r(r+r_-)} \,. 
\eea
 
 \subsubsection{At spatial infinity}
 
When $r \rightarrow \infty$, the asympotic  expansion of  the matrix $M(r)$ in \eqref{systemBCLodd} reads
\bea
M(r) = M_0 + \frac{1}{r} M_{-1} + {\cal O}\left( \frac{1}{r^2}\right) \, , \qquad
M_0 \equiv -i\left(
\begin{array}{cc}
0 & \omega^2 \\
1 & 0
\end{array}
\right) \, , \quad 
M_{-1} \equiv {2} 
\left(
\begin{array}{cc}
1 & 0 \\
-i \rs  & 0
\end{array}
\right)  \,,
\eea
where we have stopped at order $1/r$, which will be sufficient for our purpose. 
Note that the two terms in the above  expansion do not depend on $\xi$, which mean they coincide with the analogous terms in GR. This is consistent with the observation that the asymptotic behaviour of the potential (\ref{pot_bcl}) at infinity coincides with that of the RW potential \eqref{RWpotential} up to first order in $1/r$. 

Since we have already analysed the same asymptotic system in Paper I for the axial modes in Schwarzschild, we recall briefly the main result. 
Using the transformation 
\beq
\X=\tilde P \, \tilde{\X}\,, \qquad \tilde P = \begin{pmatrix} 
-1+\varpi_+ & 1 + \varpi_- \cr
1+ \varpi_+ & 1 - \varpi_- \end{pmatrix} \, , \qquad
 \varpi_\pm \equiv   \frac{\pm\,  \omega  \mass+i }{2 \omega r}  \,,
 \label{P_infty_bcl}
\eeq
we obtain the equivalent, and fully diagonalised, system
\bea
	&& \dv{\tilde{\X}}{r} = \tilde{M} \tilde{\X} \, , \quad 
	\tilde{M}(r) =  \left( \begin{array}{cc} - i \omega & 0 \\ 0 & i \omega \end{array}\right) + 
	\frac{1}{ r}\left( \begin{array}{cc} 1-i  \omega \mass & 0\\ 0 & 1+i \omega  \mass  \end{array}\right) + {\cal O}\left( \frac{1}{r^2}\right)\,. 
\eea
Direct integration yields the asymptotic solution 
\bea
\tilde{\X}(r)  \; = \; \left(1 + {\cal O}\left( {1}/{r}\right)\right) \left(
\begin{array}{c}
\co_-   \, e^{-i\omega r} r^{1- i  \omega \mass } \\
\co_+ \, e^{+i\omega r} r^{1+ i \omega \mass  }
\end{array}
\right) \, = 
\left(r + {\cal O}\left(1\right)\right) \left(
\begin{array}{c}
\co_-   \, e^{-i\omega r_*}  \\
\co_+ \, e^{+i\omega r_*}  
\end{array}
\right),
\label{AsymptoticOddGR}
\eea
where $\co_\pm$ are arbitrary constants and  we have reintroduced, in the last expression,  the variable $r_*$ 
associated with the BCL solution, defined in  \eqref{TortoiseBCL}
\footnote{The tortoise coordinate associated with the BCL solution has been computed in \eqref{TortoiseBCL} and its large $r$ expansion
reads
\bea
r_*= r+\mass \ln r - \frac{r_+^2 \ln r_+ - r_-^2 \ln r_-}{r_+ + r_-} - \frac{r_+^2 + r_-^2 - r_+ r_-}{r} + {\cal O}\left( \frac{1}{r^2}\right) \, .
\eea
When $\mass = \rs = 2m$, it coincides with the Schwarzschild tortoise coordinate \eqref{TortoiseS} $r_* = r + \rs \ln r$ up to the order
${\cal O}(1)$. Hence, one can equivalently use any of the two coordinates in the asymptotic \eqref{AsymptoticOddGR} which has been given up to ${\cal O}(1)$ as well.}.

Taking into account the time dependence $e^{-i\omega t}$ of the modes, the two components $\tilde \X_-$ (up component) and $\tilde \X_+$ (down component)  of $\tilde \X$ take  the form 
\beq
\label{BCLasymptrstar}
e^{-i\omega t}\tilde \X_{\mp}(r)= \co_\mp \left(r + {\cal O}\left( 1\right)\right) e^{-i \omega (t\pm r_*)}\,,
\eeq
where one recognises the usual  ingoing mode (associated with $\co_-$) and outgoing mode (associated with $\co_+$) at spatial infinity. The values of $\co_\pm$ can be restricted by the boundary conditions imposed on the system. For example, requiring that the mode is purely outgoing, as is the case for QNMs,  imposes $\co_-=0$.

\subsubsection{At the horizon}

We now turn to the asymptotic behaviour at the black hole horizon. Introducing the variable 
\bea
\varepsilon \equiv r - r_+\,,
\eea
 the near-horizon asymptotic expansions of the functions $A$,  $\cC$ and $\cD$ in \eqref{ACDBCL}, are given by 
\bea
\label{asympcCcD}
A = {\cal O}(\varepsilon) \, , \qquad \cC = i \left(\frac{\cC_{2}}{\varepsilon^2} + \frac{\cC_{1}}{\varepsilon} + \cC_0 \right)+ {\cal O}(\varepsilon)\, , \qquad \cD =  \frac{\cD_{1}}{\varepsilon} + \cD_0 + {\cal O}(\varepsilon) \,.
\eea
Substituting into \eqref{systemBCLodd}, we obtain 
the asymptotic expansion of the matrix $M$, 
\bea
M(\varepsilon) =  \frac{1}{\varepsilon^2} \begin{pmatrix} 0 & 0 \\ \cC_{2} & 0\end{pmatrix} +  \frac{1}{\varepsilon} \begin{pmatrix} 0 & 0 \\ \cC_{1} & \cD_{1}\end{pmatrix} +   \begin{pmatrix} {2}/{r_+} & -i \omega^2 \\ \cC_0 & \cD_0\end{pmatrix} + {\cal O}(\varepsilon) \,,
\eea
where  we will need only   the explicit expression of the coefficients $\cD_1$ and $\cC_2$, 
\bea
\cD_1=-1\,, \qquad \cC_2 = -i r_0^2 \,  \quad {\rm with}\quad r_0 \equiv r_+ \frac{\sqrt{r_+(r_+ + 2  r_-)}}{r_+ + r_-} \,.
\eea

Our system now differs from the GR analog studied in Paper I. However, the leading order term is still nilpotent, as in GR, and the resolution of the system is very similar to the analysis of Paper I.  According to the algorithm,  one first needs to  perform the transformation
\bea
\label{P1}
\X\equiv P_{(1)} \X^{(1)} \, , \qquad \text{with} \qquad P_{(1)} (\varepsilon) \equiv \begin{pmatrix} 1 & 0 \\ 0 & 1/\varepsilon\end{pmatrix} \, ,
\eea
which leads to the new system
\bea
\dv{{\X^{(1)}}}{\varepsilon} = M^{(1)} \X^{(1)} \, ,\quad M^{(1)}= \frac{1}{\varepsilon} \begin{pmatrix}0 & -i\omega^2 \\ \cC_2 & 1+\cD_1 \end{pmatrix} + {\cal O}(1)=- \frac{i}{\varepsilon} \begin{pmatrix}0 & \omega^2 \\ r_0^2 & 0 \end{pmatrix} + {\cal O}(1)\,.
\eea
The  leading term of the new matrix $M^{(1)}$ is now diagonalisable and  the system can be explicitly diagonalised via the transformation
\beq
\label{P2}
\X^{(1)}\equiv P_{(2)}\X^{(2)}\, , \qquad \text{with} \qquad P_{(2)} = \begin{pmatrix} \omega & -\omega \\ r_0 & r_0\end{pmatrix}  \, ,
\eeq
 leading to the new system
\bea
\dv{{\X^{(2)}}}{\varepsilon} = M^{(2)} \X^{(2)} \, ,\qquad M^{(2)}(\varepsilon) = \frac{i  \omega r_0}{\varepsilon} \begin{pmatrix} -1 & 0 \\ 0 & 1 \end{pmatrix} + {\cal O}(1)\, .
\eea

Finally, integrating this system yields
\bea
\label{X2}
\X^{(2)}(\varepsilon) = (1 + {\cal O}(\varepsilon)) \begin{pmatrix} \co_- \varepsilon^{- i  \omega r_0} \\ \co_+ \varepsilon^{+ i  \omega r_0}\end{pmatrix} 
=  (1 + {\cal O}(\varepsilon)) \begin{pmatrix} \co_- e^{- i  \ceta \omega r_*} \\ \co_+ e^{+ i \ceta \omega r_*}\end{pmatrix}  \,  ,
\eea
{ where $\co_\mp$ are constants and we have used the asymptotic expansion of the tortoise coordinate \eqref{TortoiseBCL} near  the horizon,
\bea
r_* = \frac{r_+^2}{r_+ + r_-} \ln \varepsilon + {\cal O}(1) = \frac{r_0}{\ceta}  \ln \varepsilon  + {\cal O}(1)\,,\qquad \ceta \equiv \frac{\sqrt{r_+ + 2 r_-}}{r_+^{1/2}}\,.
\eea
Taking into account the time dependence $e^{-i\omega t}$, one thus gets for the two components of $\X^{(2)}$
\beq
e^{-i\omega t}\X^{(2)}_{\mp}= \co_\mp e^{-i \omega (t\pm \eta r_*)}\left(1 + {\cal O}(\varepsilon)\right) \,,
\eeq
where one recognizes the ingoing and outgoing modes, propagating with the velocity $c=\eta^{-1}$, in agreement  with the expression (\ref{c_hor_bcl}) obtained in the previous section, via the Schr\"odinger-like equation.

\subsubsection{Numerical computation of the quasi-normal modes}
\label{subsubsec:numerical-QNM-odd}

A very useful application of knowing the asymptotic solutions at infinity and near the horizon is the numerical computation of  the quasi-normal modes (see e.g. the reviews \cite{Kokkotas:1999bd,Nollert:1999ji,Berti:2009kk}), as illustrated  in Paper I for  Schwarzschild black holes in General Relativity.  In the context of modified gravity, quasi-normal modes have been computed explicitly for a few solutions, such as  black holes in Einstein-Gauss-Bonnet  \cite{Blazquez-Salcedo:2016enn,Blazquez-Salcedo:2017txk,Blazquez-Salcedo:2020caw,Blazquez-Salcedo:2020rhf} or dynamical Chern-Simons gravity
\cite{Molina:2010fb}.

Let us briefly explain the principle of the computation,  based on  \cite{Jansen:2017oag},  and apply it to the BCL solution.
The asymptotic behaviour of the original metric variables $\X_1$ and $\X_2$, defined in   \eqref{X_axial}, can be deduced from the asymptotic solutions to the diagonalised systems and the transition matrices. 
At spatial infinity, we have  $\X=P \tilde{\X}$ where $P$ is given in   \eqref{P_infty_bcl}, and 
the asymptotic behaviour \eqref{AsymptoticOddGR} for $\tilde{\X}$ implies
\bea
\X_1 & = &(\ci_+\,  r^{i \omega \mass } e^{+i\omega r} -\ci_-\,  r^{-i  \omega \mass} e^{-i\omega r} ) (r + {\cal O}(1))  \, , \\
\X_2 & = & (\ci_+ \, r^{i  \omega \mass }  e^{+i\omega r}  + \ci_- \, r^{-i  \omega \mass } e^{-i\omega r} ) (r + {\cal O}(1))  \,,
\eea
where, for later convenience, we have chosen  the formulation in terms of $r$. A quasi-normal mode is characterized by purely outgoing boundary conditions at infinity, i.e.
\beq
\ci_-=0\,.
\eeq

At the horizon, the relation between the initial and final quantities is  $\X= P_{(1)} P_{(2)} \X^{(2)}$, where $ P_{(1)}$ and $P_{(2)}$ are defined in \eqref{P1} and \eqref{P2} respectively. 
The asymptotic solution (\ref{X2}) thus yields
\bea
\X_1
&= & \omega(\ch_- \varepsilon^{- i   \omega r_0} - \ch_+  \varepsilon^{+ i  \omega r_0} ) (1+ {\cal O}(\varepsilon)) \, , \\
\X_2
&=&  r_0 (\ch_- \varepsilon^{- i  \omega r_0 -1 } + \ch_+  \varepsilon^{+ i  \omega r_0  -1 } ) (1+ {\cal O}(\varepsilon))  \,.
\eea
For a quasi-normal mode, the boundary condition at the horizon must be purely ingoing, which requires 
\beq
\ch_+=0\,.
\eeq

Now we proceed as in Paper I  to compute numerically the first quasi-normal modes of the axial perturbations about the BCL black hole. 
We first introduce an ansatz for $\X_1$ and $\X_2$, which satisfies the  required boundary conditions,
\begin{equation}
	\begin{aligned}
		\X_1= e^{i\omega r} r^{1 + i  \omega\mass} \left(\frac{r-r_+}{r}\right)^{-i\omega r_0} f_1(r)\,,\quad
		\X_2 = e^{i\omega r} r^{1 + i  \omega\mass} \left(\frac{r-r_+}{r}\right)^{-1-i\omega r_0} f_2(r) \,,
	\end{aligned}
	\label{eq:ansatz-BCL-odd}
\end{equation}
 where the functions $f_1$ and $f_2$ are supposed to be regular  in the whole domain $[r_+, \infty[$ and  {\it bounded} at spatial
 infinity and at the horizon. 
To implement these regularity conditions, we change the coordinate variable by setting
\begin{equation}
	u = \frac{2r_+}{r} - 1\, \in [-1,+1] \, ,
\end{equation}
and decompose $f_1(u)$ and $f_2(u)$ onto the Chebyshev polynomials $T_n(u)$. We truncate the decomposition at a given order $N$,
hence we have
\begin{equation}
\label{decomp_Che}
	\begin{aligned}
		f_1(u) = \sum_{n=0}^N \alpha_n T_n(u) \,, \qquad
		f_2(u) = \sum_{n=0}^N \beta_n T_n(u) \, ,
	\end{aligned}
\end{equation}
where $\alpha_n$ and $\beta_n$ are the complex coefficients to be determined by the resolution of the equations of motion. 
The next step consists in reformulating the differential system \eqref{generaloddsystem} as a system of  linear algebraic equations
of the form, 
\begin{equation}
	M_{N}(\omega) V_N(\alpha_n, \beta_n) = 0 \,,
\end{equation}
where $M_{N}$ is a $2(N+1) \times 2(N+1)$ complex-valued matrix whose (finite) expansion in powers of $\omega$ reads\footnote{We use indices inside brackets to indicate the coefficients of the powers of  $\omega$ (in contrast with the coefficients in the asymptotic expansions at spatial infinity or near the horizon).}
\begin{equation}
	M_N(\omega) = M_{N[0]} + M_{N[1]} \omega + M_{N[2]} \omega^2 \,,
\end{equation}
while the $2(N+1)$-dimensional column vector  $V_N(\alpha_n, \beta_n)$ contains the coefficients of the decompositions \eqref{decomp_Che}
\begin{equation}
	{}^T V_N(\alpha_n, \beta_n) \equiv \begin{pmatrix} \alpha_0, &  \cdots, & \alpha_N, & \beta_0, & \cdots, & \beta_N \end{pmatrix} \, .
\end{equation}
Following  \cite{Jansen:2017oag}, we transform the problem of solving the previous linear system in terms of a generalised eigenvalue problem which is formulated as follows,
\begin{equation}
	\tilde{M}_{N}(\omega) \tilde{V}_N(\alpha_n,\beta_n) = 0 \,,
	\label{eq:syst-generalized-eigs-odd}
\end{equation}
where the dimensions of the matrix $\tilde{M}_{N}$ and the vector $\tilde{V}_N$ have been doubled compared to the previous (untilded) ones according to
\begin{equation}
    \tilde{M}_N =  \tilde{M}_{N[0]} + \tilde{M}_{N[1]} \omega \quad \text{and} \quad     \tilde{M}_{N[0]} = \begin{pmatrix} M_{N[0]} & M_{N[1]} \\ 0 & \mathbf{1} \end{pmatrix} \,, \quad \tilde{M}_{N[1]} = \begin{pmatrix} 0 & M_{N[2]} \\ - \mathbf{1} & 0 \end{pmatrix} \,.
\end{equation}

At this stage, it is finally possible to compute the values of $\omega$ using Mathematica or Scipy. To proceed, we computed the modes for two different values of $N$ and kept the ones that agree up to a given precision, which allows us to get rid of the spurious solutions. The first quasi-normal modes have been represented in Fig. \ref{fig:QNMs-BCL-odd}.
We have also plotted, in Fig. \ref{fig:QNMs-BCL-odd_xi}, the ``evolution" in the complex plane of the first three modes ($n=0, 1,2$) for $\ell=2,3$ when $\xi$ increases.  One observes a decrease of both the real and (absolute value of the) imaginary parts of the complex frequencies as $\xi$ increases.

\begin{figure}[!h]
 \captionsetup{singlelinecheck = false, format= hang, justification=raggedright, font=footnotesize, labelsep=space}
	\centering
	\includegraphics[height=7cm]{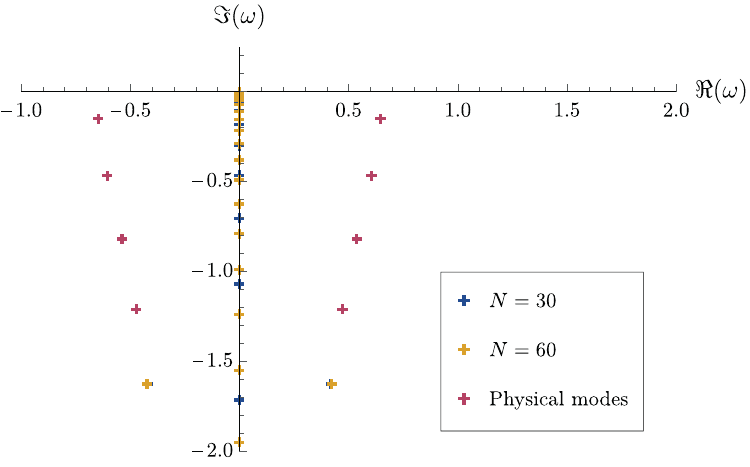}
	\caption{Quasinormal modes numerically found for $\xi = 0.5$, $\mass = 1$ and $\ell = 2$. We take $N=30$, then $N=60$, and keep the values that agree up to $10^{-3}$. The eigenvalues shown in red correspond to the physical quasinormal modes, whereas the eigenvalues visible in blue or orange correspond either to spurious modes (on the imaginary axis) or to modes that have not yet converged. We can observe that there is a symmetry about the imaginary axis. The first three modes detected are $\omega_0 = \pm 0.646 - 0.152 i$, $\omega_1 = \pm 0.605 - 0.468 i$ and $\omega_2 = \pm 0.534 - 0.819 i$.}
	\label{fig:QNMs-BCL-odd}
\end{figure}
\begin{figure}[!h]
 \captionsetup{singlelinecheck = false, format= hang, justification=raggedright, font=footnotesize, labelsep=space}
	\centering
	\includegraphics[height=7cm]{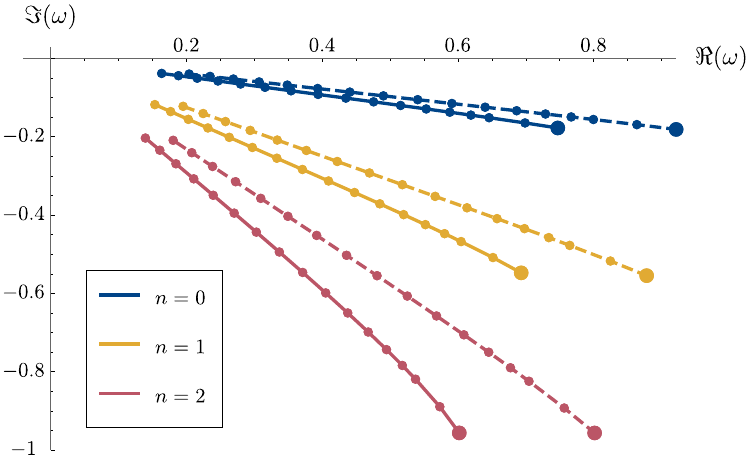}
	\caption{The first three quasinormal modes ($n=0,1,2$) for $\ell=2$ (continuous line) and $\ell=3$ (dashed line), with $\mass = 1$, when $\xi$ varies from $0$ to $50$. On each ``trajectory'',  the large dot denotes the GR mode ($\xi=0$) and the next point corresponds to $\xi=0.2$, the subsequent values of $\xi$ increasing with a constant logarithmic increment until the final value $\xi=50$.} 
	\label{fig:QNMs-BCL-odd_xi}
\end{figure}

\subsection{Stealth Schwarzschild axial perturbations}

Let us now  study the asymptotic behaviour of axial perturbations for the stealth Schwarzschild solution. As we saw in section \ref{Sec_Stealthpotential},  the dynamics of axial perturbations is now governed by the system \eqref{generaloddsystem}, 
\bea
\label{diffStealthagain}
 \dv{{\X}}{r} = M \X \, , \qquad M(r) =  \begin{pmatrix} {2}/{r}+i\omega {\cA} &- i \omega ^2 +2 i {\clam}{\cB}/{r^2}   \\ -i{ \cC} & {\cD} +i\omega\cA  \end{pmatrix} \, ,
  \eea
where the functions $\cA$, $\cB$, $\cC$ and $\cD$ are given in \eqref{ABCStealth}. Let us also recall that the constant $\zeta$ \eqref{czetadef} parametrizes  the deviation to General Relativity which is recovered in the limit $\zeta \rightarrow 0$. 

\medskip
 Following our remark, at the end of section \ref{Sec_Stealthpotential},
that the Schr\"odinger-like equation for  axial modes  is equivalent to a standard Regge-Wheeler equation, we now show that this property can be seen directly  with the first order system, via appropriate rescalings of the time and radial variables. We first perform
a time shift \eqref{changetime} with $\nu' = -\Psi$ so that $\Psi$ disappears from the  above matrix $M(r)$ in \eqref{diffStealthagain}. Then,  introducing the new
variables
\bea
 \tilde r \equiv (1 + \zeta) r \, , \qquad
 \tilde r_g \equiv (1 + \zeta) r_g \, , \qquad
 \tilde t \equiv \sqrt{1 + \zeta}\,  t\ \implies \    \tilde\omega = \omega / \sqrt{1 + \zeta} \, ,
 \label{eq:tilded-variables-rg}
\eea
one can see  that the first order differential system takes exactly the same form as in GR, namely 
\bea
 \dv{{\X}}{\tilde r} = \tilde M \X \, , \qquad \tilde M(\tilde r) =  \begin{pmatrix} {2}/{\tilde r} &- i \tilde \omega ^2 +2 i {\clam} \frac{ \tilde r - \tilde r_g}{\tilde r^3}   \\ -i \frac{\tilde r^2}{(\tilde r - \tilde r_g)^2} & -\frac{\tilde r_g}{\tilde r (\tilde r - \tilde r_g)}   \end{pmatrix} \, ,
  \eea
with  $\tilde{r}_g$ as Schwarzschild radius. 

As a consequence, the asymptotic behaviour of $Y$ is immediately  deduced from the GR results given in Paper I (section III.B).  
Both at infinity and near the horizon, the asymptotic behaviours of the  two components of $Y$ are linear combinations (with coefficients that can depend on real powers of $r$ or $\varepsilon$)  of the following outgoing and ingoing modes, 
\bea
e^{\pm i \, \tilde \omega \tilde r_*}= e^{\pm  i \, \omega  r_*}\,,   \qquad \tilde r_* \equiv \tilde r + \tilde r_g \ln(\tilde r/\tilde r_g -1) \,,
\label{eq:modes-stealth-odd-RG}
\eea
where $\tilde r_*$ corresponds to the standard tortoise coordinate in Schwarzschild (with radial coordinate $\tilde r$ and horizon $\tilde\rg$) and 
 $r_*$ is the radial coordinate introduced in  \eqref{rstar_stealth} in order to get $c(r)=1$.

 One can finally reintroduce the time dependence, taking into account the time shift $\nu$,  to obtain the asymptotic limits.
At spatial infinity, using  $\nu(r)=-\int \Psi(r) dr \approx -2\zeta \sqrt{r_s r}$, one finds 
\beq
  e^{-i\omega (t+\nu)} e^{\pm  i \, \omega  r_*} \approx e^{-i\omega (t+\nu)}e^{\pm i \omega \sqrt{1+\zeta}\, (r+(1+\zeta) r_s \ln r)}\approx
 e^{-i\omega t}e^{2i \omega\zeta \sqrt{r_s r}}e^{\pm i \omega \sqrt{1+\zeta}\, r} r^{\pm i \omega(1+\zeta)^{3/2} r_s}\,.
 \eeq
  At the horizon $r = r_g$, using  $\nu \approx  - (1+\zeta)^{3/2} \rs \ln({r}/\rg - 1) \approx -  {r}_* $, 
one gets
\begin{equation}
  e^{-i\omega (t+\nu)} e^{\pm  i \, \omega  r_*}\quad  \longrightarrow \quad e^{- i {\omega} ({t} - 2 {r}_*)} \qq{and} e^{-i {\omega} {t}} \,.
    \label{eq:axial-modes-horizon-rg}
\end{equation}
In the original coordinate system, only one mode seems to be propagating at the horizon. It is necessary to use a more appropriate time coordinate  to identify one outgoing and one ingoing mode. The above expressions could also be obtained by applying the algorithm of Paper I to the original system.

\section{Polar perturbations}
\label{section_polar_new}

We now turn to the study of polar, or  even-parity, perturbations.
We choose the same (Zerilli) gauge fixing as usually adopted in General Relativity (see e.g. Paper I for details), thus the metric perturbations are parametrised 
by four families of functions $H_{0}^{\ell m}$, $H_{1}^{\ell m}$, $H_{2}^{\ell m}$ and $K^{\ell m}$ ($\ell$ and $m$ are integers with
 $\ell \geq 0$ and $-\ell \leq m \leq \ell$) such that the non-vanishing components of the metric are
\bea
\label{eq:even-pert-Horn}
&&h_{tt} = A(r)\sum_{\ell, m} H_{0}^{\ell m}(t,r) Y_{\ell m}(\theta,\varphi), \quad
 h_{rr} = B(r)^{-1} \sum_{\ell, m} H_{2}^{\ell m}(t,r) Y_{\ell m}(\theta,\varphi) ,  \nonumber \\
&&h_{tr} = \sum_{\ell, m} H_{1}^{\ell m}(t,r) Y_{\ell m}(\theta,\varphi), \quad h_{ab} = \sum_{\ell, m} K^{\ell m}(t,r) g_{ab} Y_{\ell m}(\theta,\varphi) \, , 
\eea
where the indices $a, b$ belong to $\{\theta, \varphi\}$. 
The scalar field perturbation is parametrised by one family of functions
according to
\begin{equation}
\delta\phi =  \sum_{\ell, m} \delta\phi^{\ell m}(t, r) Y_{\ell m}(\theta, \varphi) \, .
\end{equation}
In the following we will consider only the modes  $\ell \geq 2$ (the monopole $\ell=0$ and the dipole $\ell=1$ require different gauge fixing conditions).

We will study successively the BCL and stealth Schwarzschild solutions. Essentially, we proceed as in the previous section for axial perturbations. 
The main difference is that the first order system is now four-dimensional since it contains a scalar mode and a gravitational mode, which are coupled. 
By contrast with the axial case, we have not been able to reduce the system to a 2-dimensional Schr\"odinger-like equation, so the only option available to us in this case is the 
asymptotic analysis of the first-order system. We thus use the algorithm of Paper I to obtain the behaviour of the solutions of the system near the horizon and at spatial infinity. Since the calculations are more involved than in the axial case, we have summarised the steps of the procedure in the main text and confined the details to Appendix \ref{Appendix:BCLeven}.

\subsection{BCL solution}
In the frequency domain, the linear equations of motion can be written as  a four-dimensional first-order differential system (see Appendix \ref{Appendix:BCLeven} for details)
\bea
\frac{\mathrm{d}\X}{\mathrm{d}r}=M \X \,,
 \eea
with the column vector 
\beq
\X={}^T\!(K \, \chi \, H_1 \, H_0)\,,
\eeq 
where $\chi$ corresponds to a renormalised scalar field perturbation, namely
\bea
\label{normalisedscalar}
\chi(r) \equiv \frac{\bb}{\af \sqrt{A(r)}} \delta \phi(r) \,.
\eea
The explicit form of the  square matrix $M$ can be read off from the equations of motion (see discussion in Appendix  \ref{Appendix:EoM_perts_polar} )
\bea
M=\begin{pmatrix}
 -\frac{1}{r} + \frac{\cU}{2 r^3 A} &
 \frac{\cU}{r^4}&  \frac{i(1+\lambda)}{\omega r^2} & \frac{\cV}{r^3} \\
\frac{\omega^2 r^2}{A^2}-\frac{\lambda}{A} - \frac{\mass}{2rA} +
\frac{\mass^2 \cS}{4r^4 A^2} &  -\frac{2}{r} - \frac{ \cU\cV}{2r^5 A} &
 -  \frac{i\omega r}{A}+   \frac{i (1+\lambda)\cU}{2r^3 \omega A} &- \frac{\lambda}{A} - \frac{3\cU}{2r^3 A} - \frac{\xi^2 \mass^4}{2 r^4 A}   \\
 - \frac{i \omega\cV}{r^2 A} & \frac{2i\omega}{r} -\frac{ i \omega\cU}{r^3 A} & -\frac{\cU}{r^3 A} & -  \frac{i \omega\cV}{r^2 A}  \\
 -\frac{1}{r} + \frac{\cU}{2r^3 A} &  \frac{2}{r^2} - \frac{\cU^2}{2 r^6 A} &  - \frac{i\omega}{A}+  \frac{i(1+\lambda)}{\omega r^2} &  \frac{1}{r}  - \frac{\cU}{2r^3 A} - \frac{\cU \cV}{2r^5 A}
\end{pmatrix} \, ,\label{MevenBCL}
\eea
where  we have introduced the functions
\bea
\label{defofcUcV}
\cU(r) \equiv \mass (r+\xi \mass) \, ,\qquad \cV(r) \equiv r^2 + \xi \mass^2 \, ,\qquad \cS(r) \equiv r^2 +  2\xi r (2 \mass -r) + 2 \xi^2 \mass^2\,.
\eea
We  analyse below the  asymptotic behaviours of the above system, first at spatial infinity and then  near the horizon. 

\subsubsection{At spatial infinity}
 The expansion of  the matrix $M$  in \eqref{MevenBCL} at spatial infinity is of the form
 \bea
 M(r) = r^2 M_2 + r M_1 + M_0 + \frac{1}{r} M_{-1} + {\cal O}\left( \frac{1}{r^2}\right) \, ,
 \eea
 where the matrices $M_i$ can  easily be inferred from \eqref{MevenBCL}. 
 
 The leading matrix $M_2$  contains a single non-zero entry,   $(M_2)_{21}=\omega^2$, and is thus nilpotent. To diagonalise the system, one can follow step by step the 
 algorithm presented in Paper I. Here, however, in order to shorten the procedure, we first adopt  a  ``customised'' strategy by considering  a transformation of the form
  \bea
 \X=P_{(1)} \X^{(1)} \, , \qquad 
 P_{(1)} = {\rm Diag}(r^{p_1}, r^{p_2}, r^{p_3}, r^{p_4})
 \label{firstchangeBCLeven}
 \eea
 and choosing the powers $p_i$ that simplify the system the most. With the choice 
 \beq
 p_1=0\,, \quad p_2=2\,, \quad p_3=p_4=1\,, 
\eeq
 one finds that the system becomes
  \bea
 \dv{{\X}^{(1)}}{r} = {M}^{(1)} {\X}^{(1)} \, , \quad  M^{(1)} = M^{(1)}_0 +  \frac{1}{r}  M^{(1)}_{-1}+ {\cal O}\left(\frac{1}{r^2} \right) \, ,
 \eea
 where the two matrices $ M^{(1)}_0$ and $ M^{(1)}_{-1}$ have the simple expressions
 \bea
 M^{(1)}_0 = \begin{pmatrix}
 0 & 0 & 0 & 1 \\
 -\omega^2 & 0 & i\omega  &  0 \\
  0 & -2i\omega & 0 & -i\omega \\
 0 & 0 & -i\omega& 0 
 \end{pmatrix} \, , \quad
 M^{(1)}_{-1} = \begin{pmatrix}
 -1 & -\mass& i (1+\lambda)/\omega& 0\\
  -2 \omega^2 \mass & -4 & 0 & -\lambda \\
  -i\omega& i \omega \mass& -1 &-i\omega \mass \\
 0 &-2 & -i\omega \mass & 0
 \end{pmatrix} \, .
 \label{MBCLevenatinfinity}
 \eea
 Following now  the algorithm of Paper I, two  additional steps are needed to obtain a fully diagonalised system  (up to order $r^0$), given by
 \bea
  \qquad \frac{\mathrm{d}\tilde \X}{\mathrm{d}r}=\tilde M \tilde \X \, , \qquad (\X = \tilde{P} \tilde{\X})\,, 
 \eea
where the (combined) transition matrix $\tilde{P}$ and the expansion of $\tilde M$ are given explicitly in Appendix \ref{Appendix:BCLeven}. 
 Integrating this asymptotic system yields
 \bea
 \tilde{\X}(r) =
 \begin{pmatrix}
 c_- \, r^{-i \omega\mass } \, e^{-i\omega r}  \\
 c_+ \, r^{+i  \omega \mass} \, e^{+i\omega r} \\
\frac{d_-}{r^3} \, r^{-\frac{ \omega\mass}{\sqrt{2}}} \, e^{-\sqrt{2}\omega r}  \\
\frac{d_+}{ r^3} \, r^{+\frac{\omega \mass}{\sqrt{2}}} \, e^{+\sqrt{2}\omega r} 
 \end{pmatrix}
 \left(1 + {\cal O}({1}/{r})\right) \, ,
 \label{eq:behav-inf-BCL-even}
 \eea
where $c_\pm$ and $d_\pm$ are constants.

The first two components are very similar to the components of the asymptotic solution obtained in the axial sector (see \eqref{AsymptoticOddGR}) and it is therefore natural  to identify these modes with  the usual outgoing and ingoing gravitational modes. By contrast, the last two components  have 
an  unusual form. 
If  we return to the original variables, via  the transformation \eqref{changeglobalBCLeven}, we find that the asymptotic behavior of the (renormalized) scalar perturbation $\chi$ \eqref{normalisedscalar} reads
\bea
\label{asympchiinfty}
\chi(r)= \frac{3}{2r}\left[{d_-} \, r^{-\frac{\omega\mass}{\sqrt{2}}} \, e^{-\sqrt{2}\omega r} -
d_+ \, r^{\frac{\omega\mass}{\sqrt{2}}} \, e^{\sqrt{2}\omega r} \right]  \left(1 + {\cal O}({1}/{r})\right) \,.
\eea

 The behaviour exhibited by this perturbation appears problematic, as it is associated with an effective metric which does not possess the appropriate causal structure. Indeed, the asymptotic  solution (\ref{asympchiinfty}) can be related to an equation of motion for $\tilde\chi\equiv r \chi$  of the form 
\beq
\label{elliptic_eq}
\frac{\partial^2\tilde\chi}{\partial t^2}+ \frac{\partial^2\tilde\chi}{\partial \tilde r ^2}\approx 0\,,\qquad {\rm with}\quad \tilde r=\sqrt{2}\left( r+\frac{\mass}{2}\ln r \right)\,,
\eeq
which does not  correspond to a wave equation. This non-hyperbolicity is usually associated with a ghost or gradient instability.

For a more direct -- although less rigorous --  approach to this problem, it is instructive to examine the perturbations of the scalar field on the fixed background geometry, in other words to ignore the backreaction of the scalar field perturbations on the metric. In this case, the equation of motion for the scalar field perturbation $\chi$ is of the form
\begin{equation}
	\pdv[2]{\chi}{t} + \frac12 A(r) \pdv[2]{\chi}{r} + \frac{1}{r} \left(1 + \frac{ \xi\mass^2}{2 r^2}\right) \pdv{\chi}{r} - W(r) \chi = 0 \,,
	\label{eq:chi-BCL-bg-fixed}
\end{equation}
where $W(r)$ is some potential, given explicitly in Appendix \ref{app:pert-fixed-metric}. Since $A>0$,  this equation has the structure of  an elliptic equation, similar to \eqref{elliptic_eq}.  In fact, it is even possible to show that  the asymptotic behaviour \eqref{asympchiinfty} can  be directly recovered from \eqref{eq:chi-BCL-bg-fixed}, as shown in Appendix \ref{app:pert-fixed-metric}.

 \subsubsection{Near the  horizon}
 \label{Sec:evenBCLhorizon}
 To obtain the asymptotic behaviour near the horizon,  we  define, as usual,  the small parameter $\varepsilon \equiv r - r_+$. It is then convenient to make the
 following  initial change of vector to simplify the analysis:
\bea
 \X=P_{(1)} \X^{(1)} \, , \qquad 
 P_{(1)} = 
 \begin{pmatrix}
 1 & 0 & 0 & 0 \\
 0 & 1/\varepsilon & 0 & 0 \\
 0 & 0 &  1/\varepsilon & 0 \\
 0 & 0 & 0 &  1/\varepsilon
 \end{pmatrix} \, .
 \eea
The  matrix $M^{(1)}$  associated to the system for $\X^{(1)}$ 
admits a very simple asymptotic expansion, of the form 
\bea
M^{(1)} =  \frac{1}{\varepsilon} M^{(1)}_0+ {\cal O}(1) \, ,
\eea
 where the matrix $M^{(1)}_0$ is given in (\ref{M1_0_polar_bcl_horizon}) in the Appendix \ref{Appendix:BCLeven}. 
  
  After transforming this matrix into a Jordan block form as shown in Appendix \ref{Appendix:BCLeven}, 
  one finds that  the asymptotic expansion of the modes reads
    \bea
  \label{asympstealthhor}
 \X^{(2)}(r) = \begin{pmatrix}
 c_- \varepsilon^{-i  \omega r_0} \\
 c_+ \varepsilon^{+i \omega r_0} \\
 ( a_1 \ln \varepsilon + a_2)  \sqrt{\varepsilon}\\
 a_1 \sqrt{\varepsilon}
 \end{pmatrix} (1+ {\cal O}(\varepsilon))\, ,
\label{eq:behav-horiz-BCL-even}
 \eea
where again $c_\pm$, $a_1$ and $a_2$ are constant.
The correspondence between the original vector $\X$  and $\X^{(2)}$ and the expression of the matrix $P=P_{(1)}P_{(2)}$ 
are described in the Appendix \ref{Appendix:BCLeven}.
 
The behaviour of the first two components in \eqref{asympstealthhor}  is the same as in the axial case, and one can thus identify them with  
 the ingoing and outgoing gravitational modes.  By contrast, the behaviour of  the last two components is very peculiar and is related to the presence of the scalar field degree of freedom. 
 As in the spatial infinity limit, these modes do not seem to correspond to a second-order equation respecting the usual four-dimensional causal structure, which indicates that the effective metric near the horizon, in which the perturbations propagate, is pathological.

 \subsubsection{Computation of the quasinormal modes}
In the following, we restrict ourselves to the ``gravitational'' modes, which behave asymptotically like the axial modes. We do not consider the ``scalar" modes, whose pathological behaviour probably indicates the presence of an instability, as mentioned earlier.
To compute numerically the quasi-normal modes, we extend the method of  section \ref{subsubsec:numerical-QNM-odd} to a 4-dimensional system.
At spatial infinity, we require the modes to be  purely  outgoing, while they must be purely ingoing near the horizon. This  implies the restrictions 
\beq
c_-=d_+=d_-=0\,, \qquad 
 c_+ = a_1=a_2=0\,,
 \eeq
in \eqref{eq:behav-inf-BCL-even} and \eqref{asympstealthhor}, respectively.  Taking into account these requirements, we consider the following ans\"atze for the four perturbations:
\begin{equation}
	\begin{aligned}
		H_0(r) &= e^{i\omega r} r^{1 + i  \omega\mass} \left(\frac{r-r_+}{r}\right)^{-1+i\omega r_0} f_0(r)\,,\\
		H_1(r) &= \omega e^{i\omega r} r^{1 + i \omega\mass } \left(\frac{r-r_+}{r}\right)^{-1+i\omega r_0} f_1(r)\,,\\
		K(r) &= e^{i\omega r} r^{i \omega\mass } \left(\frac{r-r_+}{r}\right)^{+i\omega r_0} f_K(r)\,,\\
		\chi(r) &= e^{i\omega r} r^{-1 + i \omega\mass } \left(\frac{r-r_+}{r}\right)^{-1+i\omega r_0} f_\chi(r)\,,\\
	\end{aligned}
	\label{eq:ansatz-BCL-even}
\end{equation}
where the functions  $f_0$, $f_1$, $f_K$ and $f_\chi$ are supposed to be bounded.

Decomposing these functions onto Chebyshev polynomials, up to some order $N$,  the differential system  with \eqref{MevenBCL} is transformed into the matricial equation
\begin{equation}
	M_{N}(\omega) \, V_{N} = 0 \, , 	\qquad {\rm with} \quad M_N(\omega) = M_{N[0]} + M_{N[1]} \omega + M_{N[2]} \omega^2 \,,
\end{equation}
where the components of the $4(N+1)$-dimensional  column vector $V_{N}$ are the components of the functions $f_0$, $f_1$, $f_K$ and $f_\chi$ on the Chebyshev basis. 
Once again, this linear system corresponds to a generalised eigenvalue problem and  the values of $\omega$ can be determined numerically. Changing the truncation order $N$ then enables us to identify the quasi-normal modes of the full system.

The first modes are represented in Fig. \ref{fig:QNMs-BCL-even-metric}. Even though the 
numerical analysis could  be further refined\footnote{We can see that the results are plagued with a lot of spurious eigenvalues caused by numerical errors, which prevents us from probing higher values of $\xi$, or higher-overtone modes. This problem comes from the higher order of the coupled system: it is made of four first-order equations, while the system for axial modes involves only two equations. In order to get accurate  estimates  of the frequencies, we need to increase the precision of the computations, and this is extremely time-consuming. This is the reason why we do not probe higher-overtone modes here. 
}, we can  already make interesting observations. First, when the parameter $\xi $ vanishes, all the modes found agree with the ones of Schwarzschild in General Relativity as expected. When $\xi$ is not vanishing and increases, the real and imaginary parts of the modes decrease compared to those of 
GR. It is interesting to note that we have obtained a continuous deformation of the classical branch of the polar modes in GR
and no other modes are detected. In other words, there is a one-to-one correspondance between the metric polar modes of the BCL black hole and the modes of the Schwarzschild black hole in GR. Hence, it seems that  imposing the metric boundary conditions, 
recalled above, on the equations of perturbations is sufficient to ensure only the metric modes are computed.

\begin{figure}[!h]
 \captionsetup{singlelinecheck = false, format= hang, justification=raggedright, font=footnotesize, labelsep=space}
	\centering
	\includegraphics[scale=0.7]{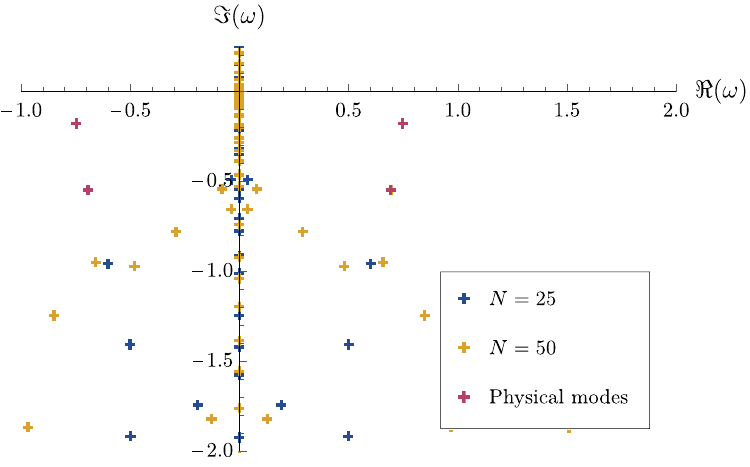}
	\caption{Quasinormal modes found for $\xi = 10^{-4}$ and $\mass = 1$. We take $N=25$, then $N=50$, and keep the values that agree up to $10^{-3}$. We can observe that there is a symmetry about the imaginary axis. Only the first two modes are detected, and they match with the Schwarzschild frequencies up to $10^{-3}$. }
	\label{fig:QNMs-BCL-even-metric}
\end{figure}

As a final but interesting  remark,  we underline that the well-known degeneracy between axial and polar modes (the so-called iso-spectrality property) in GR is lifted when one considers the BCL solution. Indeed, the polar and axial modes associated to the same overtone are different as soon as $\xi \neq 0$. This is illustrated in Fig. \ref{fig:lifting-degeneracy}. Such a feature could be used to discriminate between a GR black hole and a modified gravity black hole  in the ringdown phase of a black hole merger.
\begin{figure}[!h]
 \captionsetup{singlelinecheck = false, format= hang, justification=raggedright, font=footnotesize, labelsep=space}
	\centering
	\includegraphics[height=6cm]{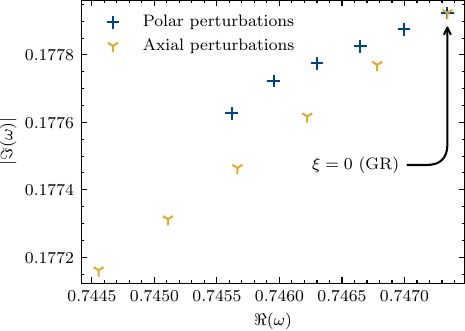}
	\caption{Tracking of the fundamental metric mode of BCL for $0 \leq \xi \leq 0.01$. The parameter $\xi$ is increased by $0.002$ between each point. We observe that the polar and axials modes are identical in the GR limit (as expected), but become different as soon as $\xi \neq 0$.}
	\label{fig:lifting-degeneracy}
\end{figure}

\subsection{Stealth Schwarzschild solution}
The asymptotic behaviour of polar perturbations for stealth Schwarzschild can be computed with the same procedure as in the BCL case, even if  it turns out to be technically more involved,  with rather tedious  calculations. Since the details are not very illuminating,  we simply give the final results in this section. 
Furthermore, to simplify the analysis, we will consider theories where  only one of the  parameters $\alpha$, $\beta$ or $\gamma$  defined in 
\eqref{paramHorn} is non zero.

In each case, we find that the asymptotic expansion of the four-dimensional column vector $\X(r)$ can be written as a linear combination
of four modes, which we will  denote $\mathfrak{g}_\pm(r)$ for the modes analogous to the axial gravitational modes and $\mathfrak{s}_\pm(r)$ for the additional modes.
There will be two families of such modes, one at spatial infinity and the other one near the horizon, which will be distinguished by the subscript $\infty$ or h, respectively. 
We give below the {\it leading order} behaviour of the modes, ignoring possible multiplicative factors that are powers of $r$  or of $\varepsilon\equiv r-\rs$ with a {\it real} exponent.

For the theories with $\beta \neq 0$ or $\gamma \neq 0$, we find the following common behaviours:
\begin{itemize}
\item at spatial infinity:
\bea
 \mathfrak{g}^\infty_\pm(r) \, \approx \, r^{\pm i \omega\rs} e^{\pm i \omega r} \,,
\eea
\item near  the horizon:
\bea
\label{Kesshor}
  \mathfrak{g}^{\rm h}_\pm(\varepsilon)  \, \approx \, \varepsilon^{\pm i  \omega \rs} \, , \qquad   \mathfrak{s}^{\rm h}_\pm(\varepsilon)  \, \approx \,  \varepsilon^{- i  \omega \rs }  \, .
 \eea
\end{itemize}
By contrast, the behaviours of the ``scalar'' modes at spatial infinity are different in the two cases:
\beq
\label{scalar_infty}
\beta \neq 0:\quad \mathfrak{s}^\infty_\pm(r)  \, \approx \,  e^{-2 i \omega \rs z} z^{\pm 2 i \sqrt{\lambda}}\,, \qquad 
\gamma \neq 0: \quad \mathfrak{s}^\infty_\pm(r)   \, \approx \,  e^{ - 2i \omega \rs z \left(  z^2/3 +  1  \right) }\,,
\eeq
where we recall that $z \equiv  \sqrt{r/\rs}$.

One observes  that, in some cases, the  $+$ and $-$ modes share exactly the same leading behaviour at spatial infinity or near the horizon. 
As a consequence, the usual distinction between ingoing and outgoing modes becomes difficult, at least at leading order, and might require to consider the next orders in the asymptotic expansion. 
 It is also worth noting that, in  the  cases  $\gamma \neq 0$ and $\beta \neq 0$, the equations for the perturbations drastically simplify, as shown in Appendix   \ref{App_K_essence} for $\gamma \neq 0$, and the asymptotic behaviour of  the scalar field can be obtained from the perturbed
conservation equation
\bea
\nabla_\mu (\delta X \, \phi^\mu) \; = \; \frac{1}{\sqrt{-g}} \partial_\mu \left( \sqrt{-g}\,  \delta X \, \phi^\mu\right) \; = \; 0 \,,
\eea
where $g_{\mu\nu}$ is the Schwarzschild metric and $\delta X$ is the perturbation of $X = \phi_\mu \phi^\mu$.  Remarkably this equation
can be solved explicitly (at least in the case $\gamma \neq 0$) and its solution reproduces exactly the asymptotic behaviour of the scalar field
 derived from  the analysis of the first order system.   

\medskip

Finally, in the case $\alpha \neq 0$, we find the following asymptotic behaviours at  spatial infinity:
\bea
 \mathfrak{g}^\infty_\pm(r) \approx e^{\pm i \omega  \, r_* + 2 i \omega \zeta \sqrt{\rs r}}  \, ,
\qquad  \mathfrak{s}^\infty_\pm(r) \approx e^{-2 i \omega \rs z} \,,
\eea
where $r_*$ is the coordinate introduced in \eqref{rstar_stealth}.
For the ``gravitational" modes,  one can clearly identify  the  ingoing and outgoing modes, and the term proportional to $\sqrt{\rs r}$ in the exponential  of $ \mathfrak{g}^{\rm h}_\pm(r)$  could be absorbed by a time redefinition of the form \eqref{changetime}.  At the horizon, the study of the asymptotic behaviour is more subtle because in that case the ``scalar" modes and the ``gravitational" modes  might  ``see" different horizons.

We will restrict our  discussion  here to  the horizon $r_g$, where the axial modes behave as in GR as we have seen. Near $r = r_g$, we find
\begin{equation}
    \mathfrak{g}^{\rm h}_+(\varepsilon)  \, \approx \, \varepsilon^{2 i \omega (1+\zeta)^{3/2} \rs }\approx e^{2i\omega r_*} \qq{and} \mathfrak{g}^{\rm h}_-(\varepsilon)  \, \approx \, 1\,, \qq{where} \varepsilon \equiv r - r_g \,.
    \label{eq:polar-modes-rg}
\end{equation}
We thus recover exactly the same behaviour as for the axial modes obtained in \eqref{eq:axial-modes-horizon-rg}.  Performing the same time shift detailed in \eqref{eq:choice-nu}, the above  modes in \eqref{eq:polar-modes-rg}  would become
\begin{equation}
    \mathfrak{g}^{\rm h}_\pm(\varepsilon) \,\approx\, e^{\pm i {\omega} {r}_*} \,,
\end{equation}
which can be interpreted as ingoing and outgoing modes.  
In summary, the polar and axial ``gravitational" modes  have similar asymptotic properties, which are more easily interpreted in the  effective metric with horizon at $r = r_g$. We leave a detailed study of the behaviour of the ``scalar'' modes for a future work. 

As mentioned in the introduction,  a detuning of the degeneracy conditions, called \enquote{scordatura}, was proposed in \cite{Motohashi:2019ymr} as a solution to the strong coupling problem  of the stealth solutions.  In order to include this type of model, the method developed here would need to be extended. Indeed, if the degeneracy conditions are not satisfied, the perturbation system contains higher order equations. They can nevertheless  be recast into a higher-dimensional first-order system, to which we can apply our method. We leave the study of these models for future work.

\section{Conclusion}
\label{section_conclusion}
In this article, we have applied the novel approach introduced in Paper I to study linear black hole perturbations in the context of
DHOST theories. The method is very generic and enables one to obtain the asymptotic behaviours of the perturbations at spatial infinity and near the black hole horizon  without reformulating their dynamics  in terms of a Schr\"odinger-like  equation. The knowledge
of these asymptotic behaviours  is essential to define and compute the quasi-normal modes,  characterised by outgoing conditions at spatial infinity and ingoing conditions at the horizon.

We have considered here  two examples of nonrotating black hole solutions within DHOST theories. The first one is the Schwarzschild stealth solution whose  geometry
is described by the usual Schwarzschild metric while the second one is the non-stealth BCL solution whose  metric is  analogous to that of Reissner-Nordstrom black hole  with the square of the electric charge effectively negative. In both cases, 
the scalar field has a non-trivial profile (but $X=\phi_\mu \phi^\mu$ is constant in the former case whereas it  depends on the radial coordinate $r$ in the latter case).

We have treated separately axial and polar perturbations. Since the scalar field perturbation is polar, axial perturbations are described by a single (gravitational) degree of freedom and are thus easier to study. In particular,
their equations of motion can be reformulated  as  a Schr\"odinger-like equation and we have found a simple method to compute explicitly   the corresponding effective potential (which depends on the choice of the radial coordinate, as the propagation speed does).
For some stealth solutions, one obtains the very peculiar property that the axial modes ``see" a Schwarzschild metric with a displaced horizon, corresponding to the disformal transformation of the original metric into the ``frame" where the propagation speed is unity.

For axial perturbations of the BCL solution, instead of computing the quasi-normal modes in the traditional way by numerically solving the Schr\"odinger-like equation, we have used the novel method of  Paper I. We have thus first computed  the asymptotic behaviours of the perturbations, at infinity and near the horizon,  from the original differential system. We have then computed, using a spectral method,  the first quasi-normal modes for BCL, finding  a deviation from general relativity. 

The study of polar perturbations and the computation of the associated quasi-normal modes is more challenging because  the scalar field and metric perturbations are now coupled and we have not found a generalised Schr\"odinger-like reformulation of the system. The only option left was thus to apply the method of Paper I,  providing  the asymptotic behaviours of the solutions at spatial infinity and near  the horizon for both types of  black holes. For the BCL solution, we have
identified two pairs of modes at the boundaries. One pair  consists of an ingoing mode and an outgoing mode,  which look similar to  the usual gravitational modes. By contrast, the other two modes, corresponding to ``scalar" modes, possess an asymptotic behaviour that  appears  pathological.  
Restricting ourselves to the ``gravitational modes",   we have computed numerically the first quasi-normal modes for the BCL solution, both for axial and polar modes. They are distinct from the GR Schwarzschild  quasi-normal modes and, as expected, the isospectrality property breaks down as the polar and the axial quasi-normal modes are now different.

 For  the stealth black hole solution, we have found that the ``gravitational" polar modes behave asymptotically as their axial counterparts. In the stealth models with $\alpha\neq 0$, their behaviour is similar to the standard GR behaviour but in a disformed Schwarzschild metric, with a different horizon and characterised by a radially-dependent time shift.
The polar modes also contain  two additional modes, due to the presence of the scalar field, for which we have computed some asymptotic limits. We leave for a future work a detailed analysis of these "scalar" modes.

This work opens a new window for the investigation of black hole perturbations in modified gravity.  The  potential 
of the new method presented  in Paper I has been illustrated here with just a couple of  examples and a rudimentary numerical treatment.  We plan to  develop it further  in the future, especially the numerical approach in order to reach a precision that would be useful for observational constraints. 
We would also like to extend our investigation to other background solutions. Note that it would  be interesting to explore the use of  the asymptotic limits as a first diagnostic tool for potential pathologies of  black hole solutions with scalar hair.

\acknowledgements{We would like to thank Eugeny Babichev, Christos Charmousis  and Gilles Esposito-Far\`ese for interesting discussions, as well as Emanuele Berti and Vitor Cardoso for instructive correspondence.}

\appendix
\section{Stealth solutions in DHOST theories}
In this Appendix,  we recall and discuss the conditions for a DHOST theory to admit stealth solutions, i.e.  solutions of modified gravity whose metric coincides with a vacuum solution of General Relativity plus a cosmological constant. 

The  main stealth solutions in {shift-symmetric} DHOST theories are described by the Schwarzschild metric and a  scalar field  of the form
\beq
\label{phiinApp}
\phi(t,r) = qt + \psi(r)\,,
\eeq
where $q$ is constant. We also assume a constant value for  $X\equiv \phi_\mu \phi^\mu$, which we denote $X_0$.

Stealth Schwarzschild  solutions can be found in  DHOST theories, with either $X_0=-q^2$ or $X_0\neq-q^2$,  provided that the functions  appearing  in the action \eqref{DHOST} 
satisfy the conditions (see Eq. (22) of \cite{Motohashi:2019sen})
\bea
&&P=P_X=Q_X=A_1+A_2=A_{1X}+A_{2X}=0\,, \label{stealth_conds}\\
&&
(X_0+q^2) A_1 =(X_0+q^2)(2A_{1X}+A_3)= 0  \quad ({\rm at} \ X=X_0) \, , \label{stealth_conds2}
\eea
where all functions are evaluated at $X=X_0$. These conditions were shown to be necessary and sufficient for the equations of motion of the metric to reduce to those of General Relativity  for static and spherical symmetric metric  \cite{Motohashi:2019sen}. 
Type Ia DHOST theories  verify  
$A_2(X)=-A_1(X)$, which implies that  the last two  conditions in \eqref{stealth_conds} are automatically satisfied. By contrast, the conditions (\ref{stealth_conds2}) are more restrictive if $X_0 + q^2 \neq 0$. These two cases were discussed in detail in  \cite{Motohashi:2019sen}. 

One can also  look for  DHOST theories such that {\it any} solution of General Relativity  (with a cosmological constant $\Lambda$), not only 
thee static spherically symmetric metric solutions, is also solution of the DHOST theory, which imposes much more stringent conditions \cite{Takahashi:2020hso}:
\bea
\label{condStealth}
P + 2 \Lambda F = 0 \, , \quad
P_X + \Lambda (4 F_X - X_0 A_{1X}) = 0 \, , \quad
Q_X=0 \, , \quad
A_1 = 0 \, \quad
A_3 + 2 A_{1X} = 0 \, ,
\eea
where  all these expressions are evaluated at $X=X_0$.
These conditions have been recently generalised to non-shift symmetric theories and to the case where matter is coupled to gravity minimally
\cite{Takahashi:2020hso}. 

\section{Background equations of motion}
\label{Appendix:EoM}
The variation of  the shift-symmetric Horndeski action \eqref{Horndeski}
yields the equations of motion  
\bea
{\cal B}_{\mu\nu}\equiv \frac{\delta S}{\delta g_{\mu\nu}}=0 \,,\qquad  {\cal B}_{\phi}\equiv \frac{\delta S }{\delta \phi}=0 \, .
\eea
Due to Bianchi identities, the equation for the scalar field is not independent from the metric equations and therefore can be ignored. 

For a   metric of the form \eqref{metric} and a scalar field profile \eqref{phi_q}, one finds that  the only non-trivial equations are given, up to a global irrelevant factor, by 
\begin{eqnarray*}
  \mathcal{B}_{tt} & \propto 
   &  \frac12 A P + q^2 P_X - \frac{A}{r^2} (-1 + B + rB') F + \frac{2q^2}{r^2} \left(1 - r B \frac{A'}{A}\right) F_X  \\ 
   && + \frac{2 A}{r^2} \dv{(r X B)}{r} F_X  
    -  \frac{4B}{r^2 A} (q^4 + q^2 X (A + rA') - r A^2 X X') F_{XX} \\
    &&+ \frac12 \left[q^2 B' \psi' + B \psi' \left(\frac{4q^2}{r} + \frac{q^2 A'}{A} - AX'\right) +2 B q^2  \psi'' \right] \, , \\
      \mathcal{B}_{tr} &  \propto &  q\psi' P_X  + \frac{q}{2r} \left( \frac{4q^2}{A} + 4 X + r X \frac{A'}{A} \right) Q_X \\
      &&- \frac{2q \psi'}{r^2} \left(B - 1+ r B \frac{A'}{A}\right) F_X 
      - \frac{4q\psi' B}{r^2}\left(\frac{q^2}{A} + X + r X \frac{A'}{A}\right) F_{XX}    \, , \\
    \mathcal{B}_{rr} & \propto      & - \frac{1}{2B} P + \frac{1}{r^2}\left( 1 - \frac{1}{B} + \frac{rA'}{A}\right) F - \frac{2q^2}{r^2 A} \left(2 - \frac{1}{B} + \frac{rA'}{A}\right) F_X \\
    &&- \frac{4X}{r^2} \left(1 - \frac{2}{B} + \frac{rA'}{A}\right)F_X 
    +  (\psi')^2 P_X + \frac{\psi'}{2r} \left( \frac{4q^2}{A} + 4 X + r X \frac{A'}{A} \right) Q_X \\
    &&- \frac{4}{r^2} B (\psi')^2 \left(\frac{q^2}{A} + X + r X \frac{A'}{A}\right) F_{XX} \, , \\
      \mathcal{B}_{\theta\theta} & \propto   &  \mathcal{B}_{\varphi\varphi}  \propto -\frac12 r^2 P + \frac12 r^2 B \psi' X' Q_X - \frac{B'r}{2} \left(\frac{2q^2}{A} + 2 X + r X \frac{A'}{A}\right) F_X \\      
&&      + \frac{r}{4} \left[2B' - r B \left(\frac{A'}{A}\right)^2 + \frac{B}{A}\left(r \frac{A'}{A}\frac{B'}{B} + 2 \frac{A'}{A} + 2r \frac{A''}{A}\right)\right] F
        \\
        &  &  - \frac{rBA'}{2A} \left(-\frac{2q^2}{A} + 2 X + r X'\right) F_X - r X' B \left(\frac{2q^2}{A} + 2 X + r X \frac{A'}{A}\right) F_{XX}
        \\& & -   \frac{rB}{2} \left[ - r X \left(\frac{A'}{A}\right)^2 + 2 \left(X' + r X \frac{A''}{A}\right)\right]   \,,
\end{eqnarray*}
where a prime denotes a derivative with respect to $r$.
 $X$ is related to $A$, $B$ and $\psi'$ by the equation 
 \bea
 X = g^{\mu\nu}\partial_\mu\phi \, \partial_\nu\phi= - \frac{q^2}{A} + B (\psi')^2 \, .
\eea
Assuming $X$ to be constant  drastically simplifies the above metric equations. For a Schwarzschild metric, the equations   admit a solution only if the stealth conditions  \eqref{stealth_conds} and \eqref{stealth_conds2} (restricted to Horndeski theories) are fulfilled.

\section{Equations of motion for the linear perturbations}

\label{Appendix:EoM_perts}

As discussed in the main text, the equations of motion for the perturbations are derived from the quadratic action $S_{\rm quad}[h_{\mu\nu},\delta \phi]$:
\bea
{\cal E}_{\mu\nu}\equiv \frac{\delta S_{\rm quad}}{\delta h_{\mu\nu}}=0 \,,\qquad  {\cal E}_{\phi}\equiv \frac{\delta S_{\rm quad}}{\delta \phi}=0 \, .
\eea
The equation $ {\cal E}_{\phi}=0$ turns out to be redundant as a consequence of Bianchi's identities, so we just need to take into account the 10 metric equations ${\cal E}_{\mu\nu}=0$.
 Furthermore, due to the spherical symmetry, the equations $\mathcal{E}_{t\varphi}$, $\mathcal{E}_{r\varphi}$ and $\mathcal{E}_{\varphi\varphi}$ are obviously equivalent  to $\mathcal{E}_{t\theta}$, $\mathcal{E}_{r\theta}$ and $\mathcal{E}_{\theta\theta}$ respectively.
 Hence, at this stage of the analysis, seven  equations at most  out of the initial ten equations are independent. We are going to see that we can reduce even more the set of independent equations. This is expected as the number of independent equations must be the same as the
 number of independent dynamical variables.

\subsection{Axial perturbations}
\label{Appendix:EoM_perts_1}
 The symmetry of the background implies that $\mathcal{E}_{tt} = 0$, $\mathcal{E}_{tr} = 0$ and $\mathcal{E}_{rr} = 0$. This leaves four non trivial independent equations for two independent functions $h_0$ and $h_1$. 
 One can thus expect that two of these equations are redundant, which is indeed the case.  
 
 First, one has  $\mathcal{E}_{\theta\theta} + 2\mathcal{E}_{\theta\varphi} = 0$. Then, one can notice that,
 out of these four equations, $\mathcal{E}_{t\theta}$ contains second-order derivatives of $h_0$ and $h_1$ while the others contain at most first order derivatives. This is an indication that $\mathcal{E}_{t\theta}$ is redundant and, as expected,  one can show that  a combination of $\mathcal{E}_{t\theta}$, $\mathcal{E}_{\theta\theta}$, $\mathcal{E}_{r\theta}$ and their derivatives vanishes. As a consequence, the dynamics of the axial perturbations is fully determined by the system consisting of the two equations
 \begin{equation}
	\mathcal{E}_{r\theta} = 0, \quad \mathcal{E}_{\theta\theta} = 0 \,,
\end{equation}
for the two variables $h_0$ and $h_1$. These two equations are first order with respect to the radial coordinate $r$, second order in $\omega$ and
are explicitly given in section \ref{Sec:eompert}.   

\subsection{Polar perturbations}
\label{Appendix:EoM_perts_polar}
Similarly to  axial perturbations, we start with seven equations of motion ${\cal E}_{\mu\nu}$ but they now depend on  five functions: $H_0$, $H_1$, $H_2$,  $K$ and $\delta \phi$. 

\subsubsection{BCL black hole perturbations}
In  the BCL case,  the equation $\mathcal{E}_{\theta \varphi}$ is  algebraic, as in GR,
 and yields 
$H_2$,
\bea
H_2 = \frac{ \mass (r+\xi \mass)}{r^3} \delta \phi + \frac{r^2 + \xi \mass^2}{r^2} H_0 \,.
\eea 
Among the remaining six equations for four independent functions,
it turns out that the four equations $\mathcal{E}_{tr}$, $\mathcal{E}_{rr}$, $\mathcal{E}_{t\theta}$ and $\mathcal{E}_{r\theta}$ are independent,  first-order with respect to  the radial coordinate and they imply the  last two ones,  $\mathcal{E}_{tt}$ and $\mathcal{E}_{\theta\theta}$.

Contrary to GR, the  remaining four equations cannot be reduced further because the system now contains two coupled degrees of freedom, the usual polar gravitational mode and the scalar mode. Hence, we obtain a system of four first order equations for the four functions $H_0$, $H_1$, $K$ and $\delta \phi$, whose explicit form is given in \eqref{MevenBCL}.

\subsubsection{Stealth black hole perturbations}
We proceed as in the previous case.  The 
equation $\mathcal{E}_{\theta\varphi}$,
\begin{equation}
\label{CombinationStealth}
 \frac{r(1 + 2q^2\alpha) - \rs}{r-\rs} H_0 - 4q^2\alpha \frac{\sqrt{r\rs}}{r-\rs} H_1 - \frac{r - (1 + 2q^2\alpha)\rs}{r-\rs} H_2 - 2q\alpha\sqrt{\frac{\rs}{r^3}} \delta\varphi = 0 \,,
\end{equation}
is algebraic and gives  $H_2$ in terms of the other functions. Once again,  the four equations ${\cal E}_{tr}$, ${\cal E}_{rr}$, ${\cal E}_{t\theta}$ and
 ${\cal E}_{r\theta}$ form a complete dynamical system for  $H_0$, $H_1$, $K$ and $\delta \phi$. It  can be written in the form
 \begin{equation}
 	M_A \dv{X}{r}  = M_B  X \, , \qquad X \equiv {}^T \begin{pmatrix} K & \delta\phi & H_1 & H_0 \end{pmatrix} \,,
 \end{equation}
where the expressions of the matrices $M_A$ and $M_B$ are quite cumbersome. To simplify, we restrict ourselves to  the case where  only $\beta\neq 0$ (and $\alpha = \gamma = \delta = 0$)  where $M_A$ and $M_B$  can be decomposed
in powers of $\omega$ according to
\begin{align}
	M_A = M_{A[0]} + M_{A[1]} \omega \, \qquad
	M_B = M_{B[0]} + M_{B[1]} \omega + M_{B[2]} \omega^2 \, ,
\end{align}
with
\begingroup
\allowdisplaybreaks
\begin{align}
	M_{A[0]} &= \begin{pmatrix}
		\frac{4 \beta  q^4 \sqrt{\rs ^3 r}}{r-\rs } & \frac{16 \beta  \rs  q^3}{r-\rs } & 0 & 8 \beta  q^4 \sqrt{\rs  r} \\
		-\frac{r \left(\rs ^2 \left(1-4 \beta  q^4\right)+2 r^2-3 \rs  r\right)}{(r-\rs )^2} & \frac{16 \beta  q^3 \sqrt{\rs ^3 r}}{(r-\rs )^2} & 0 & \frac{2 r \left(\rs  \left(4 \beta  q^4-1\right)+r\right)}{r-\rs } \\
		0 & -\frac{4 \beta  \rs  q^3}{r} & \rs -r & 0 \\
		r & -\frac{4 \beta  \rs ^{3/2} q^3}{\sqrt{r} (r-\rs )} & 0 & -r \\
	\end{pmatrix} \,,
\end{align}
\begin{align}	
	M_{A[1]} &= \begin{pmatrix}
		-2 i r^2 & -16 i \beta  q^3 \sqrt{\rs  r} & 0 & 0 \\
		0 & 0 & 0 & 0 \\
		0 & 0 & 0 & 0 \\
		0 & 0 & 0 & 0 \\
	\end{pmatrix} \,,
\end{align}
\begin{align}	
	M_{B[0]}  &= \begin{pmatrix}
		0 & -\frac{8 \beta  (\lambda +1) \rs  q^3}{r (r-\rs )} & -\frac{2 \left((\lambda +1) \rs ^2-2 \rs  r \left(\lambda +4 \beta  q^4+1\right)+(\lambda +1) r^2\right)}{(r-\rs )^2} & -\frac{16 \beta  q^4 \sqrt{\rs ^3 r}}{(r-\rs )^2} \\
		\frac{2 \lambda  r}{r-\rs } & -\frac{8 \beta  (\lambda +1) \rs ^{3/2} q^3}{\sqrt{r} (r-\rs )^2} & \frac{16 \beta  q^4 (\rs  r)^{3/2}}{(r-\rs )^3} & -\frac{2 r \left(\rs ^2 \left(\lambda +8 \beta  q^4\right)+\lambda  r^2-2 \lambda  \rs  r\right)}{(r-\rs )^3} \\
		0 & 0 & -\frac{\rs  \left(-\rs +4 \beta  q^4 r+r\right)}{r (r-\rs )} & \frac{2 \beta  q^4 \sqrt{\frac{\rs }{r}} (\rs +r)}{r-\rs } \\
		0 & 0 & -\frac{4 \beta  q^4 \sqrt{\rs ^3 r}}{(r-\rs )^2} & \frac{\rs  \left(\rs +2 \beta  \rs  q^4+r \left(2 \beta  q^4-1\right)\right)}{(r-\rs )^2} \\
	\end{pmatrix} \,,
\end{align}
\begin{align}		
	M_{B[1]} &= \begin{pmatrix}
		-\frac{i r \left(3 \rs ^2-\rs  r \left(4 \beta  q^4+5\right)+2 r^2\right)}{(r-\rs )^2} & \frac{16 i \beta  q^3 \sqrt{\rs ^3 r}}{(r-\rs )^2} & 0 & \frac{2 i r \left(\rs  \left(4 \beta  q^4-1\right)+r\right)}{r-\rs } \\
		\frac{4 i \beta  \rs ^{3/2} q^4 r^{5/2}}{(r-\rs )^3} & \frac{16 i \beta  \rs ^2 q^3 r}{(r-\rs )^3} & \frac{4 i r^2 \left(\rs  \left(4 \beta  q^4-1\right)+r\right)}{(r-\rs )^2} & -\frac{8 i \beta  q^4 \sqrt{\rs  r^5}}{(r-\rs )^2} \\
		-i r & -\frac{4 i \beta  q^3 \sqrt{\rs  r}}{r-\rs } & 0 & -i r \\
		0 & -\frac{4 i \beta  \rs  q^3 r}{(r-\rs )^2} & -\frac{i r^2}{r-\rs } & 0 \\
	\end{pmatrix} \,,
\end{align}
\begin{align}	
	M_{B[2]} &= \begin{pmatrix}
		0 & 0 & 0 & 0 \\
		-\frac{2 r^4}{(r-\rs )^2} & -\frac{16 \beta  q^3 \sqrt{\rs  r^5}}{(r-\rs )^2} & 0 & 0 \\
		0 & 0 & 0 & 0 \\
		0 & 0 & 0 & 0 \\
	\end{pmatrix} \,.
\end{align}
\endgroup
We do not write down the general equations (i.e. with generic values for $\alpha$, $\beta$, $\gamma$ and $\delta$)  which are particularly cumbersome. 

\section{Schr\"odinger-like equation from a general two-dimensional system}
\label{Sec:3DtoSchro}
In this Appendix, we consider a  two-dimensional first-order differential system of the form,
\begin{equation}
\label{App:system}
	\dv{\X}{r}  = M \X \, , \quad M(r) = M_{[0]}(r) + \omega M_{[1]}(r) + \omega^2 M_{[2]}(r) \, ,
\end{equation}
where the matrices $M_{[0]}$, $M_{[1]}$ and $M_{[2]}$ do not depend on $\omega$. Their coefficients, which are functions of $r$, will be denoted $a_n$, $b_n$, $c_n$ and $d_n$, so that 
\bea
M_{[n]}(r) = \begin{pmatrix} a_n(r) & b_n(r) \\ c_n(r) & d_n(r)\end{pmatrix}
\,.
\eea

The system admits a Schr\"odinger-like form  if one can find a new vector  $\hat{\X}$  related to $\X$ by the transformation $\X = \hat{P} \hat{\X}$, where the transition matrix $\hat P$ depends on $r$ but not on $\omega$,  leading to a system of the form 
\bea
\label{system_Sch}
\dv{\hat \X}{r}  = \hat M \hat \X \, , \qquad \text{with} \quad \hat M(r) = \frac{1}{n(r)} \begin{pmatrix} i \omega \mu(r) & 1 \\ V(r) - {\omega^2}/{c^2(r)} & i \omega \mu(r) \end{pmatrix}\, ,
\eea
where $n$, $\mu$, $V$ and $c$ are functions of $r$. In particular, $n(r)$ allows for a possible rescaling of the radial coordinate.

Using similar notations as in \eqref{App:system}, we can decompose $\hat{M}$ as
\bea
\hat{M}(r) = \hat{M}_{[0]}(r) + \omega \hat{M}_{[1]}(r) + \omega^2 \hat{M}_{[2]}(r) \,,
\eea
where the invidual matrices can be read off  from  \eqref{system_Sch} and are related to the matrices in \eqref{App:system} by
\bea
\label{conditionMhat}
\hat{M}_{[2]} = \hat{P}^{-1} M_{[2]} \hat P \, , \quad
\hat{M}_{[1]} = \hat{P}^{-1} M_{[1]} \hat P \, , \quad
\hat{M}_{[0]} = \hat{P}^{-1} M_{[0]} \hat P - \hat P^{-1} \hat P' \, ,
\eea
where $\hat{P}'$ denotes the derivative of $\hat{P}$ with respect to $r$.
One notices from \eqref{system_Sch} that $\hat{M}_{[1]}$ is proportional to the identity matrix and $\hat{M}_{[2]}$ is nilpotent. Given \eqref{conditionMhat}, this requires that  the original matrices ${M}_{[1]}$  and ${M}_{[2]}$ satisfy the same properties, respectively. This implies in particular that $\hat{M}_{[1]} = M_{[1]}$ and therefore
\beq
\frac{\mu}{n}=a_1=d_1\,.
\eeq

In the following, we will assume, for simplicity, that 
\beq
\label{M2_c2}
{M}_{[2]}= \begin{pmatrix} 0 & 0 \\ c_2 & 0 \end{pmatrix}\,.
\eeq
Indeed, since ${M}_{[2]}$ is nilpotent, it is always possible to make a transformation $X=\tilde P \tilde X$ to bring the matrix coefficient of $\omega^2$ in this form, so there is no loss of generality with the above assumption. It is then easy to check, 
 using the first relation in \eqref{conditionMhat},
 that the most general $\hat P$ that  brings ${M}_{[2]}$ of the form (\ref{M2_c2}) into  $\hat{M}_{[2]} $ corresponding to \eqref{system_Sch} is 
\bea
\label{generalP}
\hat P=     x\begin{pmatrix} 1 & 0 \\ y & z \end{pmatrix}  \qquad {\rm with} \quad z=-  c^2 n \, c_2\,, 
\eea
where $y$ and $x$ are  arbitrary (and $x \neq 0$). 

The functions $x$ and $y$ can be determined by requesting that the initial matrix $M_{[0]}$ is transformed into the requested form $\hat{M}_{[0]}$. Using the third transformation relation in \eqref{conditionMhat}, this leads to the four equations 
\bea
&&x' - (a_0 + y b_0) x=0 \, ,\label{eqsystem1} \\
&&1- b_0 n z = 0 \, , \label{eqforc}\\
&&(xy)'-(c_0 + y d_0) x +  V \frac{x z}{n} = 0 \, , \label{eqsystem2}\\
&& (x z)' - d_0 x z + \frac{xy}{n} = 0 \, . \label{eqsystem3}
\eea
The second equation, Eq.  \eqref{eqforc}, is purely algebraic and is solved by
 \beq
 \label{z}
 z=\frac{1}{n b_0}\,,
 \eeq
which can be substituted  into both \eqref{eqsystem2} and \eqref{eqsystem3}. The combination of  \eqref{eqsystem1} and \eqref{eqsystem3} then yields 
\bea
\label{sols_x_y}
x = \sqrt{b_0 n} \exp\frac{1}{2}  \left[ \int^r \mathrm{d}u ( a_0(u) + d_0(u)) \right] \,, \qquad y = \frac{1}{2 b_0} \left( d_0 - a_0 + \frac{b_0'}{b_0} + \frac{n'}{n}\right) \,,
\eea
and, finally, the expression of the potential follows from \eqref{eqsystem2},
\bea
V = n^2 b_0 \left( c_0 + y d_0 - y \frac{x'}{x}  - y' \right) \, .
\eea
Substituting the solutions \eqref{sols_x_y} for $x$ and $y$,  we obtain the simple expression
\begin{equation}
\begin{aligned}
\label{potentialintermsofmatrix}
V  =  \frac{n^2}{4} \left[ 4 b_0 c_0 + (d_0 - a_0)^2 -2 (d_0' - a_0') + 2 \frac{b_0'}{b_0} (d_0 - a_0) + 3 \left(  \frac{b_0'}{b_0}  \right)^2 +  \left(  \frac{n'}{n}  \right)^2  - 2  \left(  \frac{b_0''}{b_0}  +  \frac{n''}{n} \right) \right] \, .
\end{aligned}
\end{equation}
This potential, valid for an arbitrary choice of  radial coordinate, i.e. of $n$,  is associated with the propagation speed
\bea
\label{speedofpropagation}
c^2 = -\frac{1}{n^2 b_0 c_2  } \,,
\eea
obtained from (\ref{generalP}) and (\ref{z}).

In conclusion,  for any differential system of the form \eqref{App:system}, we have found the necessary and sufficient conditions for
it to be rewritten in a Schr\"odinger-like form: $M_{[1]}$ must be proportional to the identity matrix  and $M_{[2]}$ nilpotent.
In this case, and assuming the form (\ref{M2_c2}) for the matrix $M_{[2]}$ we have obtained explicitly  the potential  $V$  and  the propagation speed $c$, given respectively 
by   \eqref{potentialintermsofmatrix} and \eqref{speedofpropagation}.

Let us apply the above results to the system   \eqref{generaloddsystem} for axial perturbations. One must first transform the system so that the matrix coefficient of $\omega^2$ has the canonical form \eqref{M2_c2}. This can be done via the transformation $\X=\tilde P \tilde \X$ with 
\bea
\label{App_Ptilde}
\tilde P = \begin{pmatrix} 0 & 1 \\ 1 & 0 \end{pmatrix} \, , \quad \tilde c_2 = b_2 \, ,
\eea
so that the new (non vanishing) coefficients are
\bea
&&\tilde a_0 = d_0= \cD  \, , \quad
\tilde b_0 = c_0= - i \cC  \, , \quad
\tilde c_0 = b_0 = 2i \lambda \frac{\cB}{r^2} \, , \quad
\tilde d_0 = a_0=\frac{2}{r} \, , 
\\
&&
 \tilde a_1 = \tilde d_1 = a_1=i \cA \, , \quad
\tilde c_2 = b_2=-i \,.
\eea
Using  the expressions \eqref{speedofpropagation} and \eqref{potentialintermsofmatrix} with the ``tilded'' coefficients we obtain, respectively, 
 \bea
 \label{c2_Gamma}
c^2 =  \frac{1}{n^2 \cC}  
\eea
and 
\begin{equation}
\begin{aligned}
V  &=  \frac{n^2}{4} \left[ \frac{8(1+ \lambda \cB \cC) }{r^2}+ \cD^2 - \frac{4 \cD}{r} + 2 \cD' + 
\frac{2\cC'}{\cC} \left( \frac{2}{r} - \cD \right) + 3 \left( \frac{\cC'}{\cC} \right)^2 + \left(\frac{n'}{n} \right)^2
- 2  \left(\frac{\cC''}{\cC} + \frac{n''}{n} \right) \right] \, .
\end{aligned}
\label{generalVHorndeski}
\end{equation}

The radial rescaling is arbitrary and one can choose it so that the propagation speed is normalised, i.e. $c=1$. According to \eqref{c2_Gamma}, this corresponds to the choice
\beq
n_{c=1}=\frac{1}{\sqrt{\cC}}\,.
\eeq
Substituting in the general expression, the potential becomes 
\beq
\begin{aligned}
V_{c=1}  &=  \frac{1}{4\cC} \left[ \frac{8(1+ \lambda \cB \cC) }{r^2}+ \cD^2 - \frac{4 \cD}{r} + 2 \cD' + 
\frac{2\cC'}{\cC} \left( \frac{2}{r} - \cD \right) +   \frac{7\,\cC^{\prime 2}}{4\, \cC^2} -  \frac{\cC''}{\cC} \right] \, .
\end{aligned}
\label{generalV_c=1}
\eeq
This expression can  be applied in particular to the first-order system governing  polar perturbations about a Schwarzschild black hole in GR, as  recalled in  paper I.
In this case, one recovers the usual Regge-Wheeler potential.

\section{Details  on   the BCL black hole perturbations}
\label{Appendix:BCLeven}
In this Appendix, we give more details on the asymptotic analysis of polar perturbations about the BCL solution.

\subsection{At spatial infinity}
The final variable $\tilde{\X}$ (which diagonalises the dynamical system up to the order $1/r^2$ at spatial infinity) is related
 to the original variable $\X$ by the linear transformation $\X = \tilde{P} \tilde{\X} $ with
 \bea
 \label{changeglobalBCLeven}
 \tilde{P} = \begin{pmatrix}
 \ca_1 + \cb_1& \ca_1 - \cb_1 & \cc_1 + \cd_1 & \cc_1 - \cd_1 \\
 0 & 0  & \cc_2 + \cd_2 & \cc_2 - \cd_2 \\
 \ca_3 + \cb_3& \ca_3 - \cb_3 &\cc_3 + \cd_3 & \cc_3 - \cd_3 \\
 \ca_3 + \cb_3 & -\ca_3 + \cb_3 & \cc_4 + \cd_4 & \cc_4 - \cd_4
  \label{eq:chgvar-inf-BCL-even}
 \end{pmatrix} \, , 
 \eea
 where the coefficients are given by
 \bea
&& \ca_1 = -\frac{2 \lambda}{ 3 r \omega^2} \, , \quad \cb_1 = \frac{i (3 r - 2 \rs)}{3r\omega}  \, , \quad \cc_1 = \frac{27 - 10 \lambda}{12 r \omega^2} \, , \quad \cd_1 = - \frac{\sqrt{2}(12 r + 7 \rs) }{24 r \omega} \, , \nonumber \\
 && \cc_2 = \frac{\sqrt{2} (3 - 2 \lambda)r}{8 \omega} \, , \quad \cd_2 = \frac{(12 r + 7 \rs) r}{8} \, , \quad \ca_3 = \frac{3r + \rs}{3}  \, , \quad
 \cb_3 = - \frac{i \lambda}{3 \omega} \, ,\\
 && \cc_3 = \frac{i(2 \lambda - 9)}{6 \omega}  \, , \quad \cd_3 = \frac{i \sqrt{2}(11 \rs - 12 r)}{12}  \, , \quad
 \cc_4 = \frac{12 r - 5 \rs}{12}  \, , \quad \cd_4 = - \frac{\sqrt{2} (27 + 2 \lambda)}{12 \omega}  \, .   \nonumber
 \eea
As we announced, this change of variable enables us to diagonalise the system whose associated  matrix $\tilde M$ is
 \bea
 \tilde M(r) = \omega 
 \begin{pmatrix}
- i  & 0 & 0 & 0 \\
 0 & i & 0 & 0 \\
 0 & 0 & -\sqrt{2}  & 0 \\
 0 & 0 & 0 & {\sqrt{2} }  
 \end{pmatrix} +
  \frac{1}{r}  \begin{pmatrix}
 -i  \omega\mass  & 0 & 0 & 0 \\
 0 & i  \omega \mass & 0 & 0 \\
 0 & 0 & -3 - \frac{ \omega\mass}{\sqrt{2}} & 0 \\
 0 & 0 & 0 & -3 + \frac{ \omega\mass}{\sqrt{2} } 
 \end{pmatrix} + {\cal O}\left( \frac{1}{r^2}\right) ,
  \eea
 up to the order ${\cal O}(1/r^2)$. {One can easily check that the dominant term in the asymptotic expansion of $\tilde M$ is a diagonalisation of $M_0^{(1)}$ \eqref{MBCLevenatinfinity} as expected.}

\subsection{Near the horizon}
As we showed in \autoref{Sec:evenBCLhorizon}, we can make a first change of variables
\bea
 \X=P_{(1)} \X^{(1)} \, , \qquad 
 P_{(1)} = 
 \begin{pmatrix}
 1 & 0 & 0 & 0 \\
 0 & 1/\varepsilon & 0 & 0 \\
 0 & 0 &  1/\varepsilon & 0 \\
 0 & 0 & 0 &  1/\varepsilon
 \end{pmatrix} \, ,
 \eea
so that the new differential system satisfied by $\X^{(1)}$  has an associated matrix $M^{(1)}$ 
with a very simple asymptotic expansion,
\bea
M^{(1)} =  \frac{1}{\varepsilon} M^{(1)}_0+ {\cal O}(1) \, ,
\eea
 where the matrix $M^{(1)}_0$ is given by,
\bea
\label{M1_0_polar_bcl_horizon}
 M^{(1)}_0 = 
 \begin{pmatrix}
 \frac{1}{2} &  \frac{\ceta}{r_0 r_+} & i  \frac{1+\lambda}{\omega r_+^2} &  \frac{\ceta^2}{r_+} \\
  \frac{r_+^2}{4} +  \frac{\omega^2 r_0^2 r_+^2}{\ceta^2} &  \frac{\ceta^2}{2} & i \frac{1+\lambda}{2\omega} -  \frac{i \omega r_+ r_0 }{\ceta} & 
   \frac{5-\ceta^2 + 2 \ceta^4+4\lambda }{4\ceta} r_0 \\
  0 & -i\omega & 0 & -i \ceta \omega r_0 \\
  0 & - \frac{\ceta}{2 r_0} & - \frac{i\omega r_0}{\ceta} & \frac{1-\ceta^2}{2} \, .
 \end{pmatrix} \, .
 \eea
Even though the expression of $M^{(1)}$ is relatively complex,  it can be transformed into a simple Jordan block form with two Jordan blocks. 
Indeed,   we make a new change of variable $\X^{(1)}=P_{(2)} \X^{(2)}$ where $P_{(2)}$ transforms $M^{(1)}_0$ according to
 \bea
 M^{(1)}_0 = P_{(2)} \begin{pmatrix} - i \omega r_0  & 0 & 0 & 0 \\
 0 & +i \omega  r_0 & 0 & 0 \\
 0 & 0 & 1/2 & 1 \\
 0 & 0 & 0 & 1/2 
 \end{pmatrix}P_{(2)}^{-1} \, ,
 \eea
 The solution for $\X^{(2)}$ is obtained immediately and reads
  \bea
 \X^{(2)}(r) = \begin{pmatrix}
 c_- \varepsilon^{-i \omega r_0 } \\
 c_+ \varepsilon^{+i \omega r_0 } \\
 ( a_1 \ln \varepsilon + a_2)  \sqrt{\varepsilon}\\
 a_1 \sqrt{\varepsilon}
 \end{pmatrix} (1+ {\cal O}(\varepsilon))\, ,
\label{eq:behav-horiz-BCL-even}
 \eea
where $c_\pm$, $a_1$ and $a_2$ are constant.
  
 The asymptotic expansion at the horizon of the original variable $\X$ whose components
 are the metric and  scalar perturbations \eqref{MevenBCL} is obtained directly from the matrix of change of variables $P$ such that
 $\X=P \X^{(2)}$. It is given by the product $P=P_{(1)}P_{(2)}$ which reads after a direct calculation,
\bea
P=   \frac{1}{\varepsilon} \begin{pmatrix}
 -\frac{2\rho \varepsilon \left(  i \eta r_+ \omega + 1+\lambda \right)}{\omega r_+^{3/2} \Delta_1} & 
 \frac{2\rho \varepsilon \left(  i \eta r_+ \omega -1- \lambda \right) }{\omega r_+^{3/2} \Delta_2}& 
 -\frac{2\rho \varepsilon \left( (3+2\lambda) r_+ + r_- \right) }{r_+ \Delta_3}& 
 i \varepsilon \frac{ 4 (2 r_+ + 3 r_-) r_+^3 \omega^2 -(1+\lambda) \rho^2  }{r_+^2 \Delta_3} \\
 -\frac{2i \eta r_- r_+^{3/2} }{ \Delta_1 } & \frac{2i \eta r_- r_+^{3/2} }{\Delta_2} & -\frac{r_+ ( r_++2 r_-)}{\rho } 
 & \frac{i}{2\omega} \\
 -\frac{i r_+^{1/2} (\rho + 2i \eta r_+^2 \omega)}{\Delta_1 } & - \frac{i r_+^{1/2} (\rho + 2i \eta r_+^2 \omega)}{\Delta_2 } &
 0 &1\\
   1 &  1 &  1  & 0
\end{pmatrix}
\label{eq:chgvar-horiz-BCL-even}
 \eea
where  we introduced the notations 
$\rho \equiv r_+ + r_-$ and
\bea
\Delta_1 \equiv \sqrt{r_+} (2 \omega r_+^{2} + i \eta \rho) \, , \quad
\Delta_2 \equiv \sqrt{r_+} (2 \omega r_+^{2} - i \eta \rho)   \, , \quad
\Delta_3 \equiv \rho^2 + 4 \omega^2 r_+^4 \, .
\eea

\section{Polar perturbations about the stealth solution: the K-essence case}
\label{App_K_essence}

In this appendix, we study the  polar perturbations of the stealth solution in Horndeski theory where the only non-vanishing parameter among    \eqref{paramHorn} is $\gamma\neq 0$. In this case, the action \eqref{DHOST} reduces 
to the sum of the Einstein-Hilbert term supplemented with a so-called K-essence term and simply reads 
\begin{equation}
    S[\phi, g_{\mu\nu}] = \int d^4 x \sqrt{-g} \left( R+ \frac{\gamma}{2} (X + q^2)^2 \right) \,.
\end{equation}
Following the notations and the procedure we described in the paper, we can compute the corresponding polar perturbations equations about the stealth solution. As expected, they can  be cast into a form very similar to those of GR, with three first order equations 
\begin{equation}
    \label{eq:syst-gamma-deltaX}
    \begin{aligned}
        &K' - \frac{1}{r} H_0 - \frac{i (\lambda + 1)}{r^2 \omega} H_1 + \frac{2r-3\rs}{2r(r-\rs)} K = \frac{i q^2 \gamma \sqrt{r\rs}}{\omega(r-\rs)} \delta X \,,\\
        & H_1' + \frac{i r \omega}{r-\rs} H_0 + \frac{\rs}{r(r-\rs)} H_1 + \frac{i r \omega}{r-\rs} K = 0 \,,\\
        & H_0' - K' +\frac{\rs }{r(r-\rs) } H_0 + \frac{i r \omega }{r-\rs } H_1= 0 \,.
    \end{aligned}
\end{equation}
along with one algebraic relation,
\begin{equation}
\label{Kessalgebraic}
\begin{multlined}0=\left(2\lambda + \frac{3\rs}{r}\right) H_0 + \left(\frac{i(\lambda+1)\rs}{r^2\omega} - 2ir\omega\right) H_1 \\ + \left(- \frac{4r\lambda(r-\rs)+2r\rs-3\rs^2}{2r(r-\rs)} + \frac{2r^3\omega^2}{r-\rs}\right) K + \left(\frac{2q^2\gamma r^2 \rs}{r-\rs} + \frac{i q^2\gamma \sqrt{r \rs} \rs}{\omega(r-\rs)}\right) \delta X\end{multlined} \,,
\end{equation}
where we have chosen to keep explicitly  $\delta X$, the linear perturbation of $X=\phi_\mu \phi^\mu$. $\delta X$  can also be expressed in terms of $\delta \phi$, $H_0$, $H_1$ and $K$:
\begin{equation}
    \label{eq:delta-X}
    \delta X = - \frac{q^2 (\rs +r)}{r - \rs} H_0 + \frac{2 q^2 \sqrt{\rs  r}}{r - \rs} H_1+ 2 q \sqrt{\frac{\rs }{r}} \delta\phi'+\frac{2 i q r \omega  }{r - \rs}\delta\phi \,.
\end{equation}

At this stage, it is  possible to treat the system \eqref{eq:syst-gamma-deltaX} and \eqref{Kessalgebraic} in the same way we have treated the system for polar perturbations in GR (see Paper I). We first  solve the algebraic equation  \eqref{Kessalgebraic} for $H_0$ and then substitute the solution into the first two differential equations \eqref{eq:syst-gamma-deltaX}. Hence, we obtain  a system of the form
\begin{equation}
    \label{eq:syst-GR-source-deltaX}
    \dv{\X}{r} - M(r) \X = \frac{q^2\gamma \,  \delta X}{(r-\rs)(2r\lambda+ 3\rs)}\begin{pmatrix} 2r^2\rs - 2i \sqrt{r\rs} (r\lambda + \rs) / \omega \\ \rs r^2 \sqrt{r\rs} / (r-\rs) - 2 i r^4 \rs \omega / (r-\rs) \end{pmatrix} \,,
\end{equation}
where $\X \equiv {}^T (K \, H_1)$ and $M(r)$ is  the matrix entering in the dynamical  system of polar perturbations in  GR whose expression has been computed in the companion paper, 
\bea
M(r) = \frac{1}{3 \rs +2 \lambda  r}\begin{pmatrix}
	\frac{\rs  (3\rs +(\lambda -2) r) - 2 r^4 \omega ^2}{r (r-\rs ) } & \frac{2 i (\lambda +1) (\rs
   +\lambda  r)+2 i r^3 \omega ^2}{r^2 } \\
 \frac{i r \left(9\rs^2-8 \lambda  r^2+8 (\lambda -1)\rs  r\right) + 4 i r^5 \omega ^2 }{2 (r-\rs )^2 } & \frac{2 r^4 \omega ^2-\rs  (3 \rs +3 \lambda  r+r)}{r (r-\rs )} \\
 \end{pmatrix} \, .
\eea
We can therefore interpret the system \eqref{eq:syst-GR-source-deltaX} as describing the dynamics of unmodified polar perturbations in GR on which the scalar field acts like a source.

Finally, it is possible to obtain a fully decoupled equation for the perturbation $\delta X$. For that one  replaces the expressions of $H_0'$, $H_1'$ and $K'$ (computed from \eqref{eq:syst-GR-source-deltaX} or the algebraic equation) into \eqref{eq:syst-gamma-deltaX}. After a direct calculation, one obtains
\begin{equation}
    \label{eq:delta-X-gamma-decoupled}
    i r^2 \left( \sqrt{r\rs} - 2 i r^2 \omega \right)\delta X'(r) + \left(\frac32 i r \sqrt{r\rs} + r^3 \left(3 - \frac{r}{r-\rs}\right) \omega + \frac{2 i r^5}{r - \rs} \sqrt{\frac{r}{\rs}} \omega^2 \right) \delta X(r) = 0 \,,
\end{equation}
which, after some simplifications, becomes\footnote{Notice that such a decoupled equation for $\delta X$ was expected. Indeed, we can directly check that it is exactly the same as  the well-known conservation equation (for linear perturbations) in shift-symmetric theories,
\bea
\nabla_\mu \left( \sqrt{-g}\,  \delta X \, e^{-i \omega t} \, \phi^\mu \right) = \frac{1}{\sqrt{-g}} \partial_\mu \left( \sqrt{-g}\,  \delta X \, e^{-i \omega t} \, \phi^\mu\right) = 0 \, .
\eea
}
\bea
2 \sqrt{\rs} (r - r_s) r \, \delta X'(r)  + \sqrt{r_s}\left(3 (r -  r_s) + 2 i r^2 \sqrt{r/r_s}\right) \delta X(r) = 0 \,. 
\eea

The equation for $\delta X(r)$ can be solved explicitly and one finds
\bea
\delta X(r) = \frac{C}{r^{3/2}}  \left( \frac{\sqrt{r} + \sqrt{\rs}}{\sqrt{r} - \sqrt{\rs}} \right)^{i\omega \rs}\exp \left( - \frac{2}{3} i \omega (r+3 \rs) \sqrt{r/\rs} \right)  \, ,
\eea
where $C$ is an integration constant. Hence, the asymptotics of $\delta X$ are deduced immediately and one obtains,
\begin{equation}
    \delta X(r) \approx \frac{C}{z^3}\exp(- 2 i \omega z \rs (z^2/3 + 1)) (1 + \mathcal{O}(1/z)) \,, \quad z \equiv \sqrt{r/\rs} \gg 1\,,
\end{equation}
at infinity, and 
\begin{equation}
    \delta X(r) \approx D (r - \rs)^{-i\omega\rs} (1 + \mathcal{O}(r-\rs)) \, , \quad r-\rs \ll \rs \, ,
\end{equation}
near the horizon, where $D$ is a constant that can be computed trivially. 

\medskip

In order to compute the asymptotic behavior of $\delta\varphi$, we need to  solve \eqref{eq:delta-X}. But, at this stage,
it is already remarkable to observe that the asymptotic behaviour of $\delta X$  agrees with the asymptotic behaviour of $\delta \phi$ computed in \eqref{Kesshor} and \eqref{scalar_infty} from the first order system.

But, for completeness, let us consider  \eqref{eq:delta-X} which can be viewed as a first order equation for $\delta \phi$ with three sources
proportional to $H_0$, $H_1$ and $\delta X$. 
The first two can be computed from \eqref{eq:syst-GR-source-deltaX} and the algebraic equation while  the third one has just been computed above. By superposition, the general solution is a combination of three particular solutions (solutions where only one of the three sources is turned on) and one homogeneous solution.

The homogeneous equation is
\begin{equation}
    \delta\phi'(r) + \sqrt{\frac{r}{\rs}} \frac{i \omega r}{r-\rs} \delta\phi(r) = 0 \,.
\end{equation}
It can be fully integrated, and the solution is
\begin{equation}
    \delta\phi =  C \,  \left( \frac{z+1}{z-1} \right)^{i\omega \rs} \exp\left(-2i\omega z\rs(z^2/3 + 1) \right) \,, \quad z \equiv \sqrt{r/\rs} \,,
\end{equation}
where $C$ is also a constant. 
We observe that the solution of the homogeneous solution for $\delta \phi$ is almost the same as the solution for $\delta X$. They only
differ by the overall factor $r^{3/2}$. Hence, their behaviors at infinity and at the horizon are exactly the same ones  (up to some integers powers of $z$ that play no role). This means that the homogeneous solution and the particular solution associated with $\delta X$ have the same asymptotics. Moreover, the functions $H_0$ and $H_1$ have their asymptotic behavior fixed by the modified GR system \eqref{eq:syst-GR-source-deltaX}: they both behave like GR metric modes at infinity and at the horizon.

As a conclusion, $\delta\phi$ can have two different behaviors at infinity and at the horizon (or any linear combination of these two): it can either behave exactly like a metric mode, similarly to $H_0$ and $H_1$; or it can have the behaviour of $\delta X$ computed previously.

These behaviours are exactly the ones found for the decoupled modes \eqref{scalar_infty}. We understand now why the branches $n_+$ and $n_-$ were the same : the asymptotic scalar behaviour is set by $\delta X$, and $\delta X$ does not verify a second-order equation but a first-order one. A similar behaviour was found for the theory where $\alpha \neq 0$, which means that such a simplification of the equations may also exist in that case.

\section{Linear perturbations of the scalar field about a fixed background in Horndeski theories}
\label{app:pert-fixed-metric}

We consider a background solution, for the (static and spherically symmetric) metric $\overline{g}_{\mu\nu}(r)$ and the scalar field 
$\overline{\phi}(r,t)=qt + \psi(r)$, in Horndeski theories and we study the dynamics of the linear perturbations of the scalar field only $\delta \phi \equiv \phi - \overline{\phi}$ about such a background. Hence, we do not consider perturbations of the metric. As usual, we decompose
the perturbation of the scalar field onto spherical harmonics 
\begin{equation}
	\delta\phi =  \sum_{\ell, m} \delta\phi^{\ell m}(t, r) Y_{\ell m}(\theta, \varphi) \, ,
\end{equation}
and we study independently each components $ \delta\phi^{\ell m}(t, r) $. As these components do not couple at the linear order, we drop the indices $\ell,m$. Then, we consider the Fourier components of $\delta \phi$ which is equivalent to taking $\delta \phi(r,t)=\delta \phi(r) e^{-i \omega t}$ as we have done all along the paper. 

One can compute the equation satisfied by $\delta \phi(r)$ in any such  background but its general expression is too cumbersome to be written here. In the case $q = 0$, it can be extracted from the quadratic Lagrangian computed in \cite{Kobayashi:2014wsa}. Instead, we concentrate on the two background solutions we have considered in the paper, namely the BCL and the stealth Schwarzschild solutions. 

\subsection{BCL background}
When the background is the BCL metric, one shows that the differential equation satisfied by $\chi(t,r)$ (defined from $\delta\phi(t,r)$ in \eqref{normalisedscalar}) is given by
\begin{equation}
	\pdv[2]{\chi}{t} + \frac12 A(r) \pdv[2]{\chi}{r} + \frac{1}{r} \left(1 + \frac{\mass^2 \xi}{2 r^2}\right) \pdv{\chi}{r} - W(r) \chi = 0 \,,
	\label{eq:chi-BCL-bg-fixed}
\end{equation}
where $A(r)$ is the function entering into the BCL metric and
\begin{equation*}
	W(r)= \frac{1}{4r^4} \left(2r^2 (3+2\lambda) - 4r(1 + \lambda)\mass - 2(1+2\lambda) \mass^2 \xi - \frac12 \frac{(2r - \mass)^2}{A(r)}\right) .
\end{equation*}
As $A(r) >0$, one immediately sees that $\chi(r,t)$ satisfies an elliptic equation and is therefore not propagating.

We now consider the Fourier component of $\chi(t,r)$, namely $\chi(r)$, and change variables by writing
\begin{equation}
	\chi(r) = \varpi(r) \tilde{\chi}(r) \,.
\end{equation}
By setting
\begin{equation}
	\varpi(r) = \frac{1}{2rA(r)} \,,
\end{equation}
we obtain the following differential equation for $\tilde{\chi}$:
\begin{equation}
	\frac12 A(r) \tilde{\chi}'' + \left(\frac{\mass^2\xi}{2r^4} - W(r) - \omega^2\right) \tilde{\chi} = 0 \,.
	\label{eq:asymp-chi-tilde}
\end{equation}

When $r \longrightarrow +\infty$, this equation simplifies to
\begin{equation}
	\tilde{\chi}'' = 2 \omega^2 \tilde{\chi}\,,
\end{equation}
which means that the behaviour at infinity of $\chi(r)$ is given by
\begin{equation}
	\chi(r) = \frac{1}{2r} \left(b_1 e^{\sqrt{2} \omega r} + b_2 e^{-\sqrt{2} \omega r}\right)  \,,
\end{equation}
where $b_1$ and $b_2$ are integration constants. This agrees with the asymptotic behaviour found for the scalar mode in \eqref{asympchiinfty}. Therefore, it seems that the asymptotic behaviour of the scalar perturbation when the metric is fixed coincides with the asymptotic behavior of the scalar part of the polar modes.

In order to confirm this intuition, we study \eqref{eq:asymp-chi-tilde} when $r \longrightarrow r_+$. The resulting equation is
\begin{equation}
	\tilde{\chi}'' + \frac{1}{4(r-r_+)^2} \tilde{\chi} = 0 \,,
\end{equation}
and the general solution corresponds to
\begin{equation}
	\chi(r) = \frac{1}{\sqrt{r - r_+}} \left(b_1 + b_2 \ln(r-r_+)\right) \,,
\end{equation}
where $b_1$ and $b_2$ are integration constants. We observe that this result is also fully consistent with the asymptotic analysis in \eqref{eq:behav-horiz-BCL-even}.

\subsection{Stealth background}
A similar analysis can be make when  the background is the stealth Schwarzschild solution. For simplicity, we  distinguish again the three  
cases where the only non-vanishing parameter is $\gamma \neq 0$, $\beta \neq 0$ or $\alpha \neq 0$. 

When  $\gamma \neq 0$, the equation for $\delta \phi$ is given by,
\bea
\delta \phi'' + \frac{\rs (r- \rs) + 2i (\rs r^5)^{1/2}}{r \rs} \delta \phi' - \frac{\omega \left[ 5i(r \rs)^{3/2} - 3i (r^5 \rs)^{1/2} + 2 \omega r^4 \right]}{2 r \rs (r- \rs)^2} \delta \phi = 0 \, .
\eea
We introduce a new field $\varphi$ defined by
\begin{equation}
	\delta\phi = \kappa(r) \varphi(r) \,,
\end{equation}
where $\kappa(r)$ is chosen to eliminate the first-order derivative in the differential equation. This can be achieved with
\bea
\kappa(r) =\exp \left[- 2 i \omega \sqrt{r/\rs} (r + 3 \rs) \right]  \left( \frac{\sqrt{r/\rs} +1}{\sqrt{r/\rs} -1 }\right)^{i \omega \rs} \, ,
\eea 
and then $\varphi$ is solution of the second order equation
\bea
4 r^2 \varphi'' + \varphi = 0 \, ,
\eea
which can be solved immediately to get
\bea
\varphi(r) = a_1 \sqrt{r} + a_2 \sqrt{r} \ln r \, ,
\eea
where $a_1$ and $a_2$ are integration constants. 
We also notice that we recover the asymptotic behaviours of the scalar mode obtained in \eqref{Kesshor} and \eqref{scalar_infty}.

The case where $\beta \neq 0$ is treated in exactly the same way. Taking now
\bea
\kappa(r) =\exp \left[- 2 i \omega \sqrt{r \rs} \right]  \left( \frac{\sqrt{r/\rs} +1}{\sqrt{r/\rs} -1 }\right)^{i \omega \rs} \, ,
\eea
we show that the field $\varphi$ satisfies the equation
\bea
4 r^2 \varphi'' + (4 \lambda +1) \varphi = 0 \, ,
\eea
which, again, can be solved immediately
\bea
\varphi(r) = \sqrt{r} \left( a_+ r^{i \sqrt{\lambda}} + a_- r^{-i \sqrt{\lambda}} \right) \, ,
\eea
where $a_\pm$ are constants. We find again that the perturbation is not propagating. Furthermore, these results agree with the full asymptotic analysis of the solutions of the polar system. 

Finally, in the case  $\alpha \neq 0$,  the equation satisfied by $\delta \phi$
at linear order disappears, since the quadratic Lagrangian for $\delta\phi$ is a total derivative.

\bibliographystyle{utphys}
\bibliography{biblio_QNM}

\end{document}